\documentclass[english,journal=jpccck]{achemso}
\usepackage[T1]{fontenc}
\usepackage[latin9]{inputenc}
\usepackage{geometry}
\usepackage{color}
\usepackage{amsbsy}
\usepackage{amstext}
\usepackage{graphicx}
\usepackage{subscript}

\makeatletter

\title{Pariser-Parr-Pople Model based Configuration-Interaction Study of
Linear Optical Absorption in Lower-Symmetry Polycyclic Aromatic Hydrocarbon
Molecules}

\author{Pritam Bhattacharyya}

\email{pritambhattacharyya01@gmail.com}

\affiliation{Department of Physics, Indian Institute of Technology Bombay, Powai,
Mumbai 400076, India}

\author{Deepak Kumar Rai}

\email{deepakrai@phy.iitb.ac.in}

\affiliation{Department of Physics, Indian Institute of Technology Bombay, Powai,
Mumbai 400076, India}

\affiliation{Present Address: Department of Chemistry, Southern University of
Science and Technology, Shenzhen, 518000, China}

\author{Alok Shukla}

\email{shukla@phy.iitb.ac.in}

\affiliation{Department of Physics, Indian Institute of Technology Bombay, Powai,
Mumbai 400076, India}

\providecommand{\tabularnewline}{\\}

\@ifundefined{showcaptionsetup}{}{%
 \PassOptionsToPackage{caption=false}{subfig}}
\usepackage{subfig}
\makeatother

\usepackage{babel}
\begin{document}
\begin{abstract}
The electronic and optical properties of various polycyclic aromatic
hydrocarbons (PAHs) with lower symmetry, namely, benzo{[}ghi{]}perylene
(C$_{22}$H$_{12}$), benzo{[}a{]}coronene (C$_{28}$H$_{14}$), naphtho{[}2,3a{]}coronene
(C$_{32}$H$_{16}$), anthra{[}2,3a{]}coronene (C$_{36}$H$_{18}$)
and naphtho{[}8,1,2-abc{]}coronene (C$_{30}$H$_{14}$) were investigated.
For the purpose, we performed electron-correlated calculations using
screened, and standard parameters in the $\pi$-electron Pariser-Parr-Pople
(PPP) Hamiltonian, and the correlation effects were included, both
for ground and excited states, using MRSDCI methodology. PPP model
Hamiltonian includes long-range Coulomb interactions, which increase
the accuracy of our calculations. The results of our calculations
predict that, with the increasing sizes of the coronene derivatives,
optical spectra are red shifted, and the optical gaps decrease. In
each spectrum, the first peak representing the optical gap is of moderate
intensity, while the more intense peaks appear at higher energies.
Our computed spectra are in good agreement with the available experimental
data. For the purpose of comparison, we also performed first-principles
time-dependent density-functional theory (TDDFT) calculations of the
optical gaps of these molecules using Gaussian basis functions, and
found that they yielded values lower than our CI results.
\end{abstract}

\section{Introduction}

Nowadays, $\pi$-conjugated molecules are used for manufacturing immensely
effective as well as low-cost electronic devices such as organic thin-film
(or field-effect) transistors (OTFTs or OFETs),\cite{OTFT_1,OFET_1,OFET_2,OFET_3}
solar cells,\cite{solar_1,solar_2} and light-emitting diodes (LEDs).\cite{LED_1,LED_2,LED_3,LED_4}
Polycyclic aromatic hydrocarbons (PAHs) are a class of $\pi$-conjugated
molecules consisting of multiple aromatic rings, found to exist almost
everywhere in the universe. A considerable percentage of carbon in
the universe is present in the form of PAHs. From a technical point
of view, this species of hydrocarbons is advantageous to society in
several ways, but they are also carcinogenic to humans as well as
other living beings. PAH molecules and their isomers exhibit unique
properties and have high optical sensitivity. Therefore, to utilize
this class of molecules in technological applications, a thorough
investigation of their electronic structure and related properties
is needed.

In an earlier work involving our group,\cite{Mazumdar_et_al} electronic
structure and optical properties of coronene and related molecules
with relatively high D\textsubscript{6h} point group symmetry were
studied. In this work, our aim is to study {\small{}the lower symmetry
(C$_{2v}$}\textsubscript{}{\small{} or lower)}\textcolor{red}{{}
}derivatives of coronene so as to understand the role which symmetry
plays in determining the electronic and optical properties of PAHs.
For the purpose, we employ a Pariser-Parr-Pople (PPP) model\cite{PPP_Pople,PPP_Pariser_Parr}
based configuration-interaction (CI) methodology, established in several
of our earlier works.\cite{Aryanpour_Shukla,PhysRevB.65.125204Shukla65,PhysRevB.69.165218Shukla69,Himanshu,himanshu-triplet,Priya_Sony,dkr1,Tista1,Tista2,Tista3}
In particular, we study the optical properties of benzo{[}ghi{]}perylene
(C\textsubscript{22}H\textsubscript{12}), benzo{[}a{]}coronene (C\textsubscript{28}H\textsubscript{14}),
naphtho{[}2,3a{]}coronene (C\textsubscript{32}H\textsubscript{16}),
anthra{[}2,3a{]}coronene (C\textsubscript{36}H\textsubscript{18}),
and naphtho{[}8,1,2-abc{]}coronene (C\textsubscript{30}H\textsubscript{14}).
Several groups have studied the{\small{}se molecules} experimentally.
Khan measured the photo-absorption spectra of several cations of coronene
and its derivatives, including benzo{[}a{]}coronene and naphtho{[}2,3a{]}coronene.\cite{khan_et_al}
Bagley \emph{et al}. reported the optical absorption spectra of six-
to nine-ring PAHs including benzo{[}a{]}coronene, and naphtho{[}8,1,2-abc{]}coronene.\cite{Bagley_el_al}
The ultra-violet spectra of many large PAHs, including benzo{[}a{]}coronene,
were studied by Fetzer \emph{et al}.\cite{Fetzer_el_al} Fluorescence
emission spectra of {\small{}all the coronene derivatives}, considered
in this work, were experimentally studied by Acree \emph{et al}.\cite{Acree_et_al}
Given the fact that no previous theoretical calculations of optical
properties of these molecules exist, our work is timely. These molecules
are also interesting from another point of view; they can be seen
as finite graphene fragments with hydrogen-passivated edges. The PAH
molecules consist of aromatic rings of carbon atoms, in which the
edge carbon atoms are passivated with the hydrogen atoms. One can
imagine obtaining such molecules by cutting graphene sheets in those
shapes, in a hydrogen-rich environment. Therefore, in physics community,
such molecules have been studied extensively under the name ``finite
graphene fragments'', or graphene quantum dots.\cite{Tista1,Tista2,Tista3} 

For the purpose of benchmarking, and comparing different computational
approaches, we also performed first-principles time-dependent density
functional theory (TDDFT) calculations of the optical gaps of these
molecules using Gaussian basis functions. We find that the values
of the optical gaps obtained from the TDDFT calculations are always
lower than those obtained from our PPP model based CI calculations.

The remainder of this paper is organized as follows. In the next section,
we discuss the theoretical methodology adopted in this work. We follow
this by presenting and discussing our calculated optical absorption
spectra for various molecules, and compare our results to experiments,
wherever possible. Finally,  we present our conclusions.

\section{Theoretical approach and Computational details}

\label{sec:theory}

\subsection{Geometry}

\label{subsec:geometry}

\begin{figure}[h]
\begin{centering}
\subfloat[{Benzo{[}ghi{]}perylene}]{\begin{centering}
\includegraphics[scale=0.22]{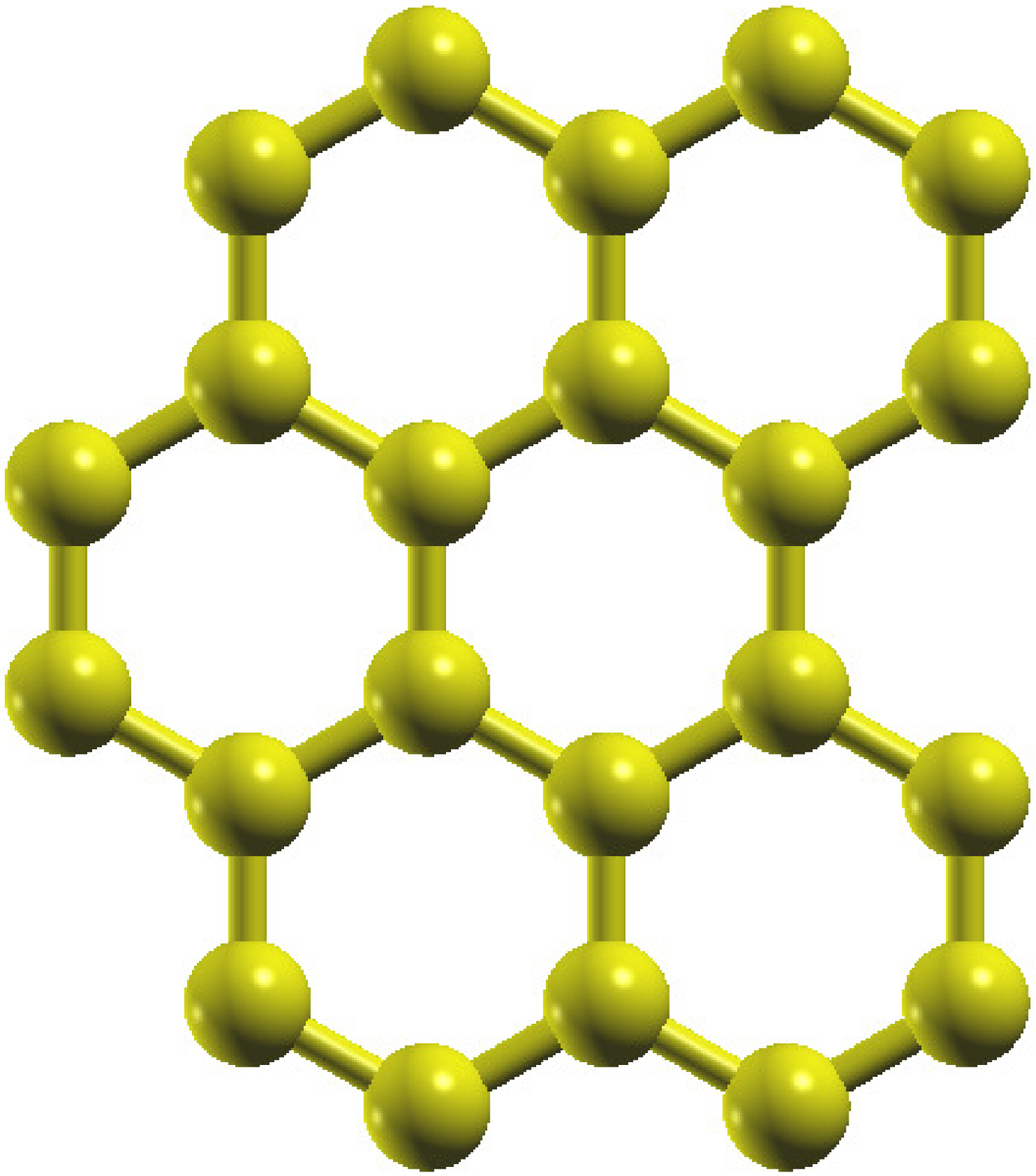}
\par\end{centering}
}~~\subfloat[{Benzo{[}a{]}coronene}]{\begin{centering}
\includegraphics[scale=0.25]{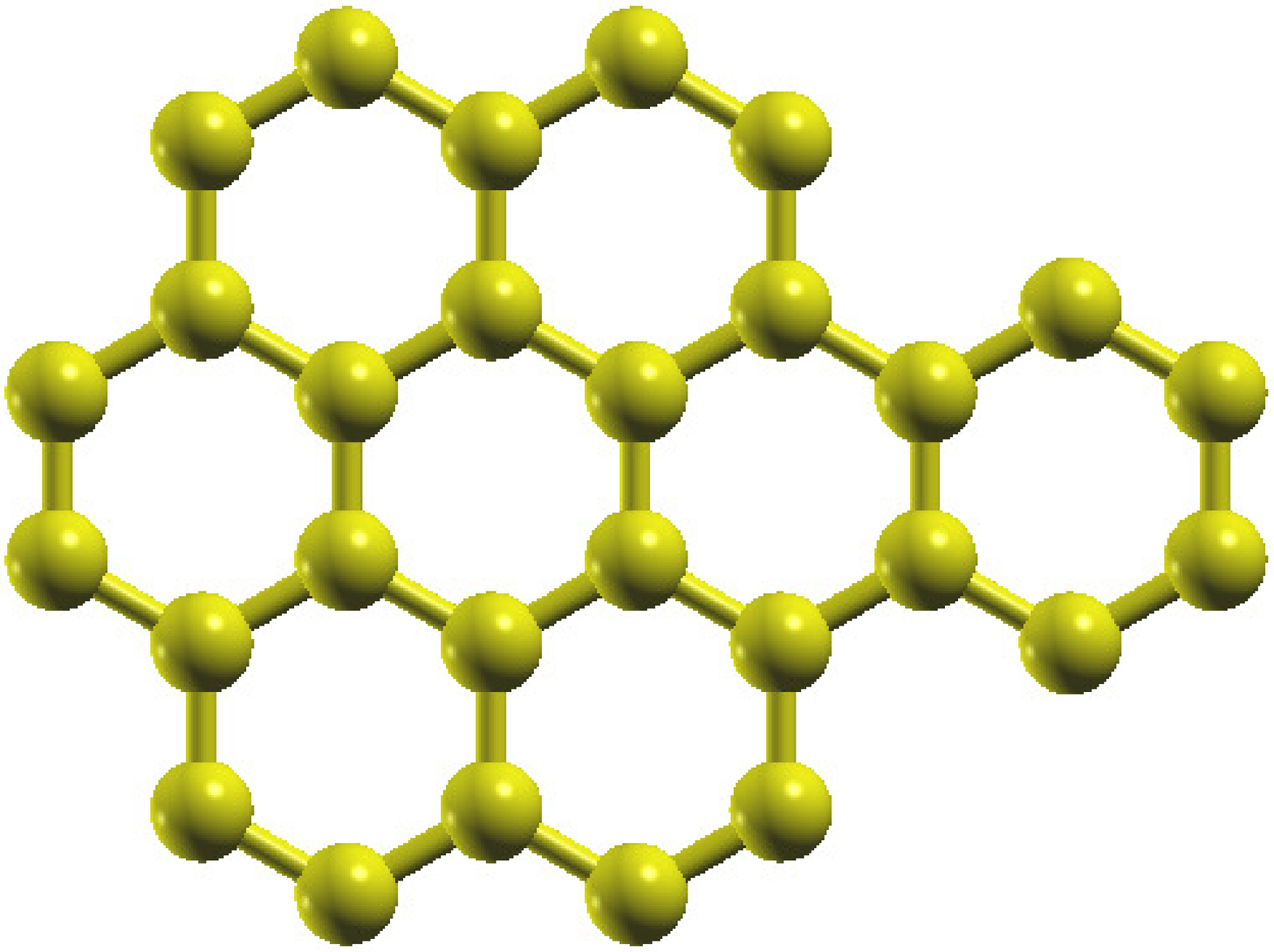}
\par\end{centering}
}~~\subfloat[{Naphtho{[}2,3a{]}coronene}]{\begin{centering}
\includegraphics[scale=0.3]{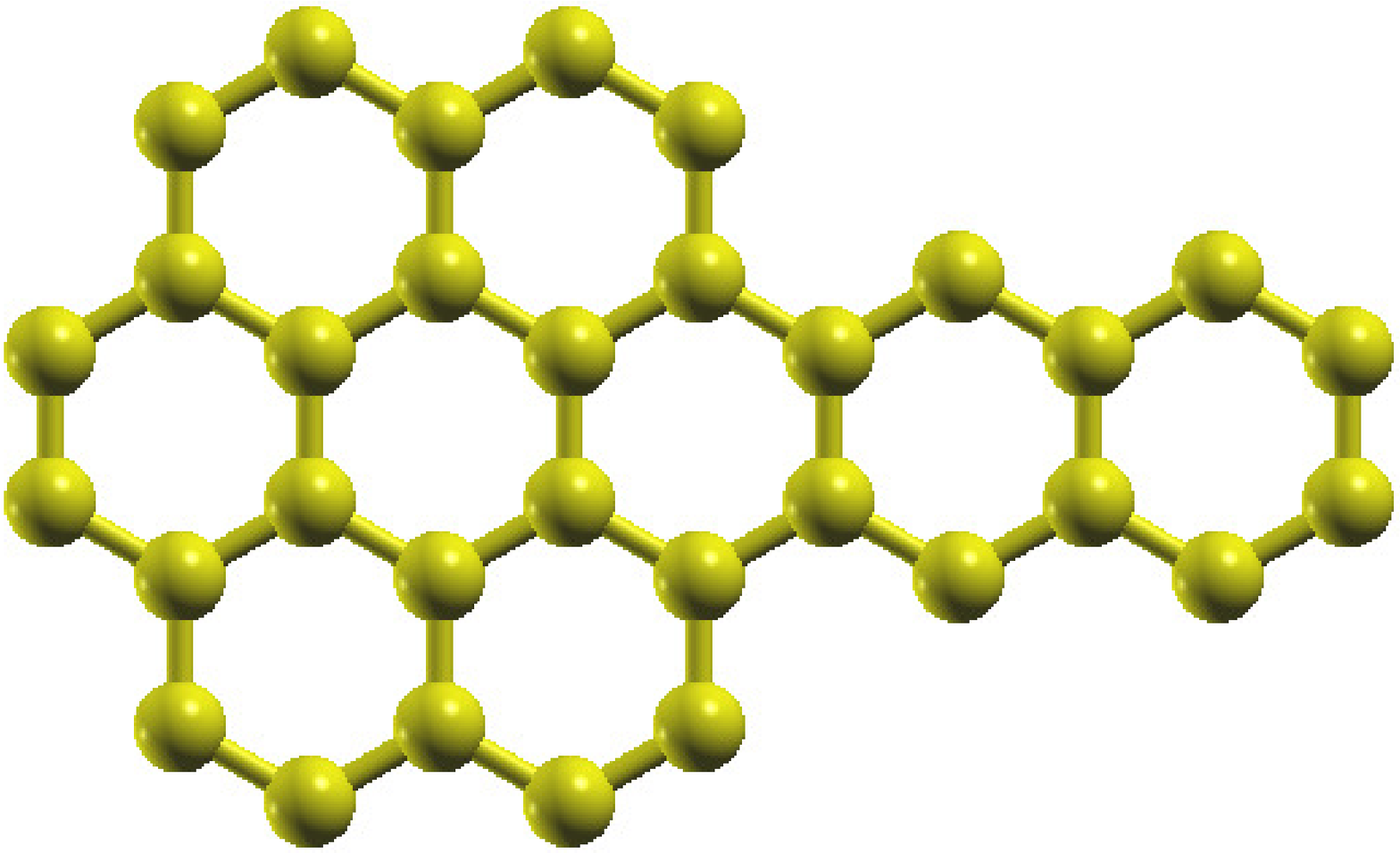}
\par\end{centering}
}\includegraphics[scale=0.4]{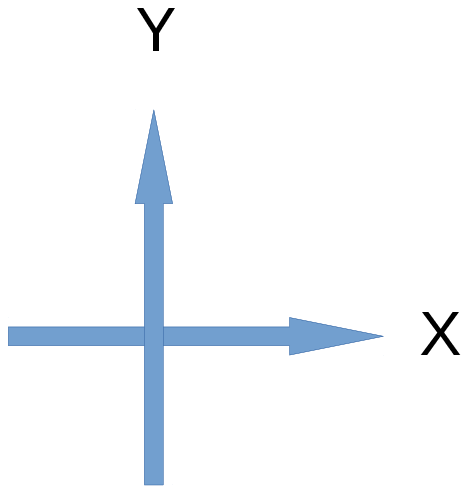}
\par\end{centering}
\begin{centering}
\subfloat[{Anthra{[}2,3a{]}coronene}]{\begin{centering}
\includegraphics[scale=0.35]{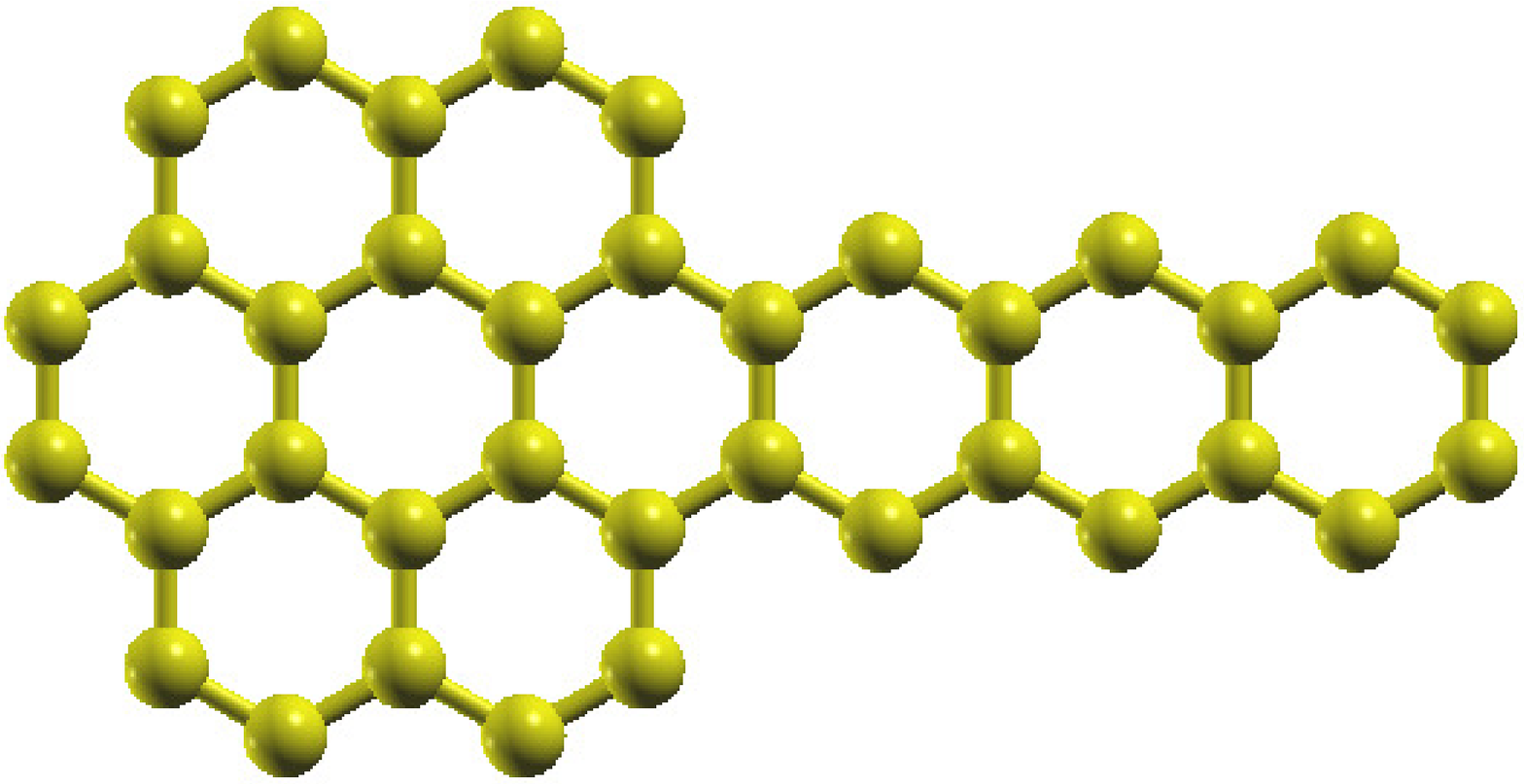}
\par\end{centering}
}~~~~~~\subfloat[{Naphtho{[}8,1,2-abc{]}coronene}]{\begin{centering}
\includegraphics[scale=0.25]{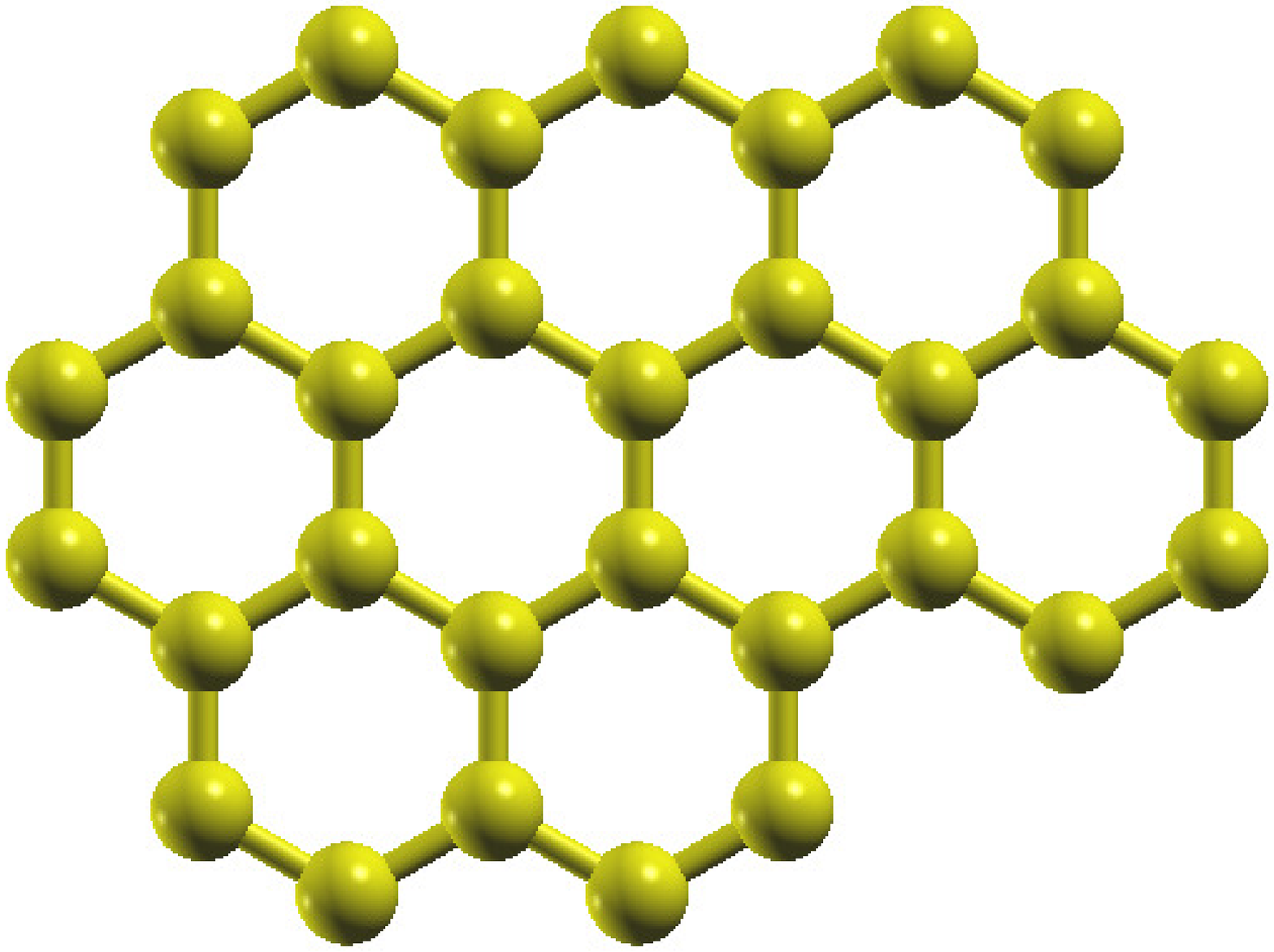}
\par\end{centering}
}
\par\end{centering}
\caption{Schematic diagrams of coronene derivatives considered in this work.
All the molecules are assumed to lie in the in $xy$-plane. The yellow
dots represent the carbon atoms, and the C-C bond lengths and bond
angles are assumed to be 1.4 \AA~ and 120$^{\circ}$, respectively.
\label{fig:geometry}}
\end{figure}

The molecules considered in this work are shown in Fig. \ref{fig:geometry}.
It is assumed that all the molecules lie in the $xy$-plane, with
uniform bond lengths and bond angles of 1.4 \AA,~ and 120$^{\circ}$,
respectively. All the molecules belong to the C\textsubscript{2v}
point group, along with a closed-shell $^{1}A_{1}$ electronic ground
state, except naphtho{[}8,1,2-abc{]}coronene, which has C\textsubscript{s}
point group, and $A^{\prime}$ ground state. According to the dipole
selection rule, the one-photon excited states have $^{1}A_{1}$ (x-polarized)
and $^{1}B_{2}$ (y-polarized) symmetries for the C\textsubscript{2v}
molecules, while for the C\textsubscript{s} molecule they have $^{1}A^{\prime}$
(xy polarized) symmetry. In this work, we only consider the excited
states corresponding to photons polarized in the plane of the molecule.

\subsection{Pariser-Parr-Pople (PPP) model Hamiltonian}

Calculations on the $\pi$-conjugated molecules considered in this
paper were performed using a semi-empirical approach, based upon PPP
model Hamiltonian,\cite{PPP_Pople,PPP_Pariser_Parr} which can be
written as

\begin{equation}
H_{PPP}=-\sum_{i,j,\sigma}t_{ij}(c_{i\sigma}^{\dagger}c_{j\sigma}+c_{j\sigma}^{\dagger}c_{i\sigma})+U\sum_{i}n_{i\uparrow}n_{i\downarrow}+\sum_{i<j}V_{ij}(n_{i}-1)(n_{j}-1)\text{.}\label{eq:H-ppp-2}
\end{equation}

In the above equation, $c_{i\sigma}^{\dagger}$ ($c_{i\sigma}$) is
the creation (annihilation) operator, \emph{i.e.}, it creates (annihilates)
a $\pi$-electron with spin $\sigma$, localized on $i^{th}$ carbon
atom. $n_{i\sigma}=c_{i\sigma}^{\dagger}c_{i\sigma}$ indicates the
total number of $\pi$-electrons with spin $\sigma$, whereas $n_{i}=\sum_{\sigma}n_{i\sigma}=\sum_{\sigma}c_{i\sigma}^{\dagger}c_{i\sigma}$
denotes the total number of $\pi$-electrons on $i^{th}$ carbon atom.
In the second and the third terms of Eq. (\ref{eq:H-ppp-2}), $U$
and $V_{ij}$ represents the on-site and long-range Coulomb interactions,
respectively. $t_{ij}$ denotes the one-electron hopping matrix element,
which in this work, has been restricted to nearest neighbors only,
with the value t\textsubscript{0} = 2.4 eV, in agreement with our
previous works on $\pi$-conjugated systems, such as conjugated polymers,\cite{PhysRevB.65.125204Shukla65,PhysRevB.69.165218Shukla69}
polycyclic aromatic hydrocarbons,\cite{Himanshu,himanshu-triplet,Aryanpour_Shukla,Priya_Sony}
and graphene quantum dots.\cite{dkr1,Tista1}

To parameterize the Coulomb interactions, we used the Ohno relationship,\cite{Ohno1964}

\begin{equation}
V_{i,j}=U/\kappa_{i,j}(1+0.6117R_{i,j}^{2})^{1/2}\mbox{,}\label{eq:Ohno}
\end{equation}

where $U$ denotes on-site electron-electron repulsion term as discussed
above, $\kappa_{i,j}$ indicates the dielectric constant of the system,
using which we can include the screening effects, and $R_{i,j}$ is
the distance between the $i^{th}$ and $j^{th}$ carbon atoms. In
this paper, we computed the optical spectra using two types of Coulomb
parameters: (a) screened parameters\cite{Chandross} {[}U = 8.0 eV,
$\kappa_{i,j}$= 2.0 (i $\neq$ j) and $\kappa_{i,i}$ = 1.0{]}, and
(b) standard parameters\cite{Ohno1964} {[}U = 11.13 eV, and $\kappa_{i,j}$=
1.0{]}. We have observed that our earlier calculations performed using
the screened parameters were in better agreement with the experimental
results, as compared to the standard parameter based ones.\cite{PhysRevB.69.165218Shukla69,Himanshu}

\subsection{Optical absorption spectrum}

For computing the optical absorption spectrum of an electronic system,
we need to obtain a good representation of its ground and excited
state wave functions. To that end, we performed the calculations using
the multi-reference singles-doubles configuration-interaction (MRSDCI)
methodology, as implemented in the computer program MELD.\cite{MELD}
For the purpose, first, we transformed the Hamiltonian from the site
representation to the molecular orbital (MO) representation, which
was achieved by performing mean-field restricted Hartree-Fock (RHF)
calculations using a code based on the PPP model, developed in our
group.\cite{SONY2010} Then, a singles-doubles CI (SDCI) calculation
was performed both for the ground state, and excited states, by employing
the transformed Hamiltonian and choosing a correct single reference
wave function. The computed excited state wave functions were used
to calculate the optical absorption spectrum at the SDCI level of
theory. Next, the excited state wave functions contributing to the
various peaks of optical spectra obtained using SDCI calculations
were used as reference states for the MRSDCI calculations. Again the
many-body wave functions of the excited states contributing to the
various peaks of optical spectra, obtained using MRSDCI calculation,
were analyzed and used as references for the next level of MRSDCI
calculation with a lower cutoff of the contributing coefficients of
the wave functions. This process is iterative and was continued until
the computed optical spectra were converged with the previous one
within an acceptable tolerance. In this study, we found the final
cutoff value of 0.05 on the magnitude of contributing coefficients
of the wave functions to be sufficient for achieving convergence.
This implies that all the configurations whose coefficients in the
MRSDCI wave function were more than 0.05 in magnitude, were included
in the list of reference configurations. This is a very stringent
cutoff, and fairly adequate for achieving convergence. In our group,
the MRSDCI approach as described above, has been used extensively
to study a variety of molecules and clusters.\cite{Pritam_jpca_et_al,Pritam_jpcs_et_al,dkr1,Aryanpour_Shukla,PhysRevB.65.125204Shukla65,PhysRevB.69.165218Shukla69,Shinde_PCCP,Tista1,Himanshu,himanshu-triplet}

For benzo{[}ghi{]}perylene, which is the smallest molecule considered
in this study, we employed the quadruples configuration-interaction
(QCI) approach to calculate the ground and the excited state wave
functions as well as the optical absorption spectra. The QCI-level
absorption spectra were compared to those computed using MRSDCI approach,
and very good agreement was obtained. Because QCI calculations are
normally more accurate as compared to MRSDCI ones, the good agreement
between the two sets of results validates the accuracy of MRSDCI method.
This is important because QCI calculations for the larger molecules
are not computationally feasible, and, therefore, only MRSDCI calculations
were performed for them.

The wave functions of the ground and the excited states thus obtained
are used to compute the electric dipole matrix elements, as well as
the optical absorption spectra $\sigma(\omega)$, by employing the
following formula

\begin{equation}
\sigma(\omega)=4\pi\alpha\sum_{i}\frac{\omega_{i0}|\langle i|\boldsymbol{\hat{e}.r}|0\rangle|^{2}\gamma^{2}}{(\omega_{i0}-\omega)^{2}+\gamma^{2}},\label{eq:sigma}
\end{equation}

where $|0\rangle$ denotes the ground state wave function, $|i\rangle$
is the wave function of the $i$-th excited state, $\omega$, $\hat{{\bf e}}$,
${\bf r}$, $\alpha$, respectively, represent the frequency of the
incident light, polarization direction of the incident light, the
position operator, and the fine structure constant. Furthermore, $\omega_{i0}$
is the energy difference (in frequency units) between the ground state
($|0\rangle$), and the $i$-th excited state ($|i\rangle$), while
$\gamma$ is the uniform lined width associated with each excited
state energy level.

In Eq. \ref{eq:sigma}, the summation over $i$ indicates a sum over
an infinite number of excited states, which, we restrict for practical
reasons to excited states with excitation energies up to 10 eV.

\subsection{Time-dependent density functional theory framework}

We also performed first-principles time-dependent density functional
theory (TDDFT) calculations to obtain the optical gap of these molecules,
so as to validate our PPP-model based results. For the purpose, we
employed a hybrid exchange-correlation functional B3LYP,\cite{Becke_TDDFT,Lee_Yang_Parr}
and a localized Gaussian basis-set, 6-31+G(d), as implemented in GAUSSIAN16
package.\cite{g16} The B3LYP/6-31+G(d) combination can exhibit excellent
comparison with the experimental , as demonstrated by Mocci et al.\cite{Mocci_et_al}
recently, for the parent molecule coronene, and discussed by us later.
Therefore, we have employed the identical methodology {[}B3LYP/6-31+G(d){]}
to study the electronic and optical properties of the PAHs considered
in this work.

\section{Results and Discussion }

\label{sec:results}

In this section, we present and analyze the computed linear optical
absorption spectra obtained using CI approach, and PPP model (PPP-CI
approach, in short) for the five PAH molecules considered in this
work. As mentioned earlier, for benzo{[}ghi{]}perylene, in addition
to QCI calculations, MRSDCI calculations were also performed using
the screened parameters in the PPP model, and the calculated absorption
spectra are compared in Fig. S1 of the Supporting Information. We
note that the peak locations of the QCI spectra are somewhat blue-shifted
as compared to the MRSDCI spectra, except for the first peak, whose
position is almost the same in both the spectra. In spite of these
small quantitative differences, the two spectra agree with each other
quite well, qualitatively. This implies that the MRSDCI methodology
is reliable, and, therefore, calculations for the four larger coronene
derivatives performed using that approach are trustworthy.

To emphasize the large-scale nature of our CI calculations, we present
the total number of spin-adapted configurations (N\textsubscript{total}),
\emph{i.e.,} the dimension of the CI matrix in Table \ref{tab:spin-adapted}
for all the symmetries of each molecule. The large numbers of spin-adapted
configurations considered in these calculations indicate that the
electron-correlation effects have been included adequately. For the
case of the smallest molecule Benzo{[}ghi{]}perylene, the total number
of configurations employed in the MRSDCI calculations was close to
one million, which is less than half of those in the QCI calculations.
In spite of that, good qualitative and quantitative agreement between
the two sets of calculations implies that MRSDCI approach is computationally
efficient, and accurate.

\begin{table}[H]
\caption{N\protect\textsubscript{total} represents the total number of spin-adapted
configurations included in the CI calculations for each symmetry of
coronene derivatives studied in this work. The superscripts 'a' and
'b' indicate the value of N\protect\textsubscript{total} employed
in the PPP calculations based on the screened, and the standard parameters,
respectively.\label{tab:spin-adapted} QCI method was employed only
for benzo{[}ghi{]}perylene, while for all other cases MRSDCI approach
was used.}

\centering{}%
\begin{tabular}{ccccccccc}
\hline 
 &  &  &  &  &  &  &  & \tabularnewline
Molecule &  & Point group &  & Symmetry &  & N$^{a}$\textsubscript{total} &  & N$^{b}$\textsubscript{total}\tabularnewline
 &  &  &  &  &  &  &  & \tabularnewline
\hline 
\hline 
 &  &  &  &  &  &  &  & \tabularnewline
Benzo{[}ghi{]}perylene  &  & C\textsubscript{2v} &  & A\textsubscript{1} &  & 2003907 &  & 2003907\tabularnewline
(C\textsubscript{22}H\textsubscript{12}) &  &  &  & B\textsubscript{2} &  & 3418335 &  & 3418335\tabularnewline
 &  &  &  &  &  &  &  & \tabularnewline
Benzo{[}a{]}coronene &  & C\textsubscript{2v} &  & A\textsubscript{1} &  & 1107236 &  & 1359626\tabularnewline
(C\textsubscript{28}H\textsubscript{14}) &  &  &  & B\textsubscript{2} &  & 688351 &  & 1598428\tabularnewline
 &  &  &  &  &  &  &  & \tabularnewline
Naphtho{[}2,3a{]}coronene &  & C\textsubscript{2v} &  & A\textsubscript{1} &  & 1252070 &  & 2611234\tabularnewline
(C\textsubscript{32}H\textsubscript{16}) &  &  &  & B\textsubscript{2} &  & 2018406 &  & 2622562\tabularnewline
 &  &  &  &  &  &  &  & \tabularnewline
Anthra{[}2,3a{]}coronene &  & C\textsubscript{2v} &  & A\textsubscript{1} &  & 3541980 &  & 4778775\tabularnewline
(C\textsubscript{36}H\textsubscript{18}) &  &  &  & B\textsubscript{2} &  & 4034925 &  & 6002917\tabularnewline
 &  &  &  &  &  &  &  & \tabularnewline
Naphtho{[}8,1,2-abc{]}coronene &  & C\textsubscript{1} &  & A\textsubscript{} &  & 2054952 &  & 4200367\tabularnewline
(C\textsubscript{30}H\textsubscript{14}) &  &  &  &  &  &  &  & \tabularnewline
\hline 
\end{tabular}
\end{table}

\subsection{Optical gap}

In order to understand the influence of electron-correlation effects
in a quantitative manner, in Table \ref{tab:optical-gap} we present
the results on the HOMO-LUMO gap of these molecules using the independent-electron
approaches, namely, the tight-binding (TB) model, and the restricted
Hartree-Fock (RHF) approach. Note that when we set U=0 (\emph{i.e.},
no electron-electron interactions) in the PPP Hamiltonian (see Eqs.
\ref{eq:H-ppp-2} and \ref{eq:Ohno}), we obtain the tight-binding
model. The same table also contains the results on optical gaps of
these molecules obtained using PPP model, and the CI approach. Optical
gaps obtained from electron-correlated calculations are counterparts
of the HOMO-LUMO gaps of one-electron theory.

\textcolor{red}{}
\begin{table}[H]
\caption{Optical gaps of five PAH molecules calculated using tight-binding
(TB), and PPP models, at various levels of theory. For one-electron
theories (TB, PPP-RHF), HOMO-LUMO (H-L) gap is interpreted as the
optical gap. For benzo{[}ghi{]}perylene reported results are at the
QCI level, while for the rest of the molecules they are at the MRSDCI
level. In the last column, experimental values of the optical gaps
(where available), are presented.\label{tab:optical-gap}}

\centering{}%
\begin{tabular}{ccccccccc}
\hline 
 & H-L gap &  & H-L gap &  & Optical gap &  & Optical gap & Optical gap\tabularnewline
Molecule & (in eV) &  & (in eV) &  & (in eV) &  & (in eV) & (in eV)\tabularnewline
 & TB &  & PPP-RHF &  & PPP-CI &  & TDDFT & experiment\tabularnewline
\cline{4-4} \cline{6-6} 
 &  &  & %
\begin{tabular}{ccc}
scr & \hspace{0.3cm} & std\tabularnewline
\end{tabular} &  & %
\begin{tabular}{ccc}
scr & \hspace{0.3cm} & std\tabularnewline
\end{tabular} &  &  & \tabularnewline
\hline 
\begin{tabular}{c}
Benzo{[}ghi{]}perylene\tabularnewline
(C\textsubscript{22}H\textsubscript{12})\tabularnewline
\tabularnewline
Benzo{[}a{]}coronene\tabularnewline
(C\textsubscript{28}H\textsubscript{14})\tabularnewline
\tabularnewline
Naphtho{[}2,3a{]}coronene\tabularnewline
(C\textsubscript{32}H\textsubscript{16})\tabularnewline
\tabularnewline
Anthra{[}2,3a{]}coronene\tabularnewline
(C\textsubscript{36}H\textsubscript{18})\tabularnewline
\tabularnewline
Naphtho{[}8,1,2-abc{]}coronene\tabularnewline
(C\textsubscript{30}H\textsubscript{14})\tabularnewline
\end{tabular} & %
\begin{tabular}{c}
2.11\tabularnewline
\tabularnewline
\tabularnewline
2.25\tabularnewline
\tabularnewline
\tabularnewline
1.89\tabularnewline
\tabularnewline
\tabularnewline
1.49\tabularnewline
\tabularnewline
\tabularnewline
1.83\tabularnewline
\tabularnewline
\end{tabular} &  & %
\begin{tabular}{ccc}
3.85 &  & 7.13\tabularnewline
 &  & \tabularnewline
 &  & \tabularnewline
3.95 &  & 7.13\tabularnewline
 &  & \tabularnewline
 &  & \tabularnewline
3.57 &  & 6.71\tabularnewline
 &  & \tabularnewline
 &  & \tabularnewline
3.14 &  & 6.23\tabularnewline
 &  & \tabularnewline
 &  & \tabularnewline
3.43 &  & 6.48\tabularnewline
 &  & \tabularnewline
\end{tabular} &  & %
\begin{tabular}{ccc}
3.39 &  & 3.54\tabularnewline
 &  & \tabularnewline
 &  & \tabularnewline
3.42 &  & 3.44\tabularnewline
 &  & \tabularnewline
 &  & \tabularnewline
2.93 &  & 3.32\tabularnewline
 &  & \tabularnewline
 &  & \tabularnewline
2.70 &  & 3.10\tabularnewline
 &  & \tabularnewline
 &  & \tabularnewline
2.95 &  & 3.23\tabularnewline
 &  & \tabularnewline
\end{tabular} &  & %
\begin{tabular}{c}
3.19\tabularnewline
\tabularnewline
\tabularnewline
3.13\tabularnewline
\tabularnewline
\tabularnewline
2.83\tabularnewline
\tabularnewline
\tabularnewline
2.35\tabularnewline
\tabularnewline
\tabularnewline
2.92\tabularnewline
\tabularnewline
\end{tabular} & %
\begin{tabular}{c}
-\tabularnewline
\tabularnewline
\tabularnewline
3.38\tabularnewline
\tabularnewline
\tabularnewline
-\tabularnewline
\tabularnewline
\tabularnewline
-\tabularnewline
\tabularnewline
\tabularnewline
3.02\tabularnewline
\tabularnewline
\end{tabular}\tabularnewline
\hline 
\end{tabular}
\end{table}

From Table \ref{tab:optical-gap} it is obvious that:\textcolor{red}{{}
}(a) for C\textsubscript{2v} symmetry coronene derivatives, the gaps
decrease with the increasing size, independently of the models and
the methods used in this work, (b) in case of TB model, the computed
gaps are smaller than the results obtained using PPP model, (c) the
gaps obtained using PPP-RHF level of theory with the standard parameters
are much larger as compared to the other computed results, (d) the
optical gaps, calculated by employing PPP-CI level of theory, are
significantly red-shifted as compared to the gaps obtained from PPP-RHF
level of theory, specially with the standard parameter calculations.
For screened parameter calculations, the correlation-induced shifts
are small, (e) the results of screened and standard parameters, obtained
using PPP-MRSDCI level of theory are in good quantitative agreement
with each for the smallest molecule, but differ somewhat for the larger
ones

When we compare our optical gaps computed using the PPP-CI approach
with those computed using the TDDFT method, we note that in all the
cases TDDFT values are smaller than the PPP-CI values (see Table \ref{tab:optical-gap}).
Although, for the case of naphtho{[}8,1,2-abc{]}coronene, our PPP-CI
band gaps computed using the screened parameters are only very slightly
(0.03 eV) larger than the TDDFT value, however, for other molecules,
the disagreement is quite significant. We hope that future measurements
of band gaps of those molecules, for which no experimental data is
available as of now, can resolve these differences. Regarding the
comparison of the compute time for the PPP-CI and the TDDFT approaches,
it depends on the level of CI approximation employed. It is well-known
that TDDFT approach is similar to CI-singles (CIS) in that both methods
account for 1 particle-1 hole excitations. For the largest molecule,
i.e., anthra{[}2,3a{]}coronene, TDDFT approach took approximately
5 hours on our cluster, while CIS calculations finish in a matter
of seconds. However, PPP-MRSDCI calculations for the same molecule
took longer than a week, because they employ a large number of configurations,
some of which have 4 particle- 4 hole character with respect to the
Hartree-Fock meanfield.

\subsubsection{Comparison with coronene (C\protect\textsubscript{24}H\protect\textsubscript{12})}

In a work involving our group, experimental measurements of the linear
and non-linear optical absorption of coronene molecule (D\textsubscript{6h}
symmetry) were performed, supported theoretically by calculations
based on PPP-CI methodology. Measurements of linear absorption found
the optical gap near 3.55 eV, characterized by very weak intensity,
while the most intense peak was found to be at 4.1 eV.\cite{Mazumdar_et_al}
Within the framework of first-principles time-dependent density functional
theory (TDDFT), Mocci et. al.\cite{Mocci_et_al} recently computed
the location of the first and the most intense peak of this molecule
to be 4.09 eV, in very good agreement with our calculated value of
4.20 eV,\cite{Mazumdar_et_al} thus, validating our PPP model-based
methodology. Our PPP-model based theory correctly predicted the most
intense peak, but was unable to predict the onset of weak absorption
at 3.55 eV, because the excited states near its location were found
to be dipole forbidden due to electron-hole symmetry selection rules.\cite{Mazumdar_et_al}
In the case of lower symmetry PAH molecules studied here, again, the
optical gap is characterized by weak intensity. However, the corresponding
optical transition is dipole allowed for these molecules, due to their
lower symmetries. As far as quantitative comparison is concerned,
for all the molecules studied here, the optical gap is lower than
3.55 eV measured for coronene.\cite{Mazumdar_et_al}

\subsection{Linear optical absorption spectrum}

In this section, we present and discuss the calculated optical absorption
spectra of the PAH molecules, considered in this work. The calculations
were performed using the CI approach, and the spectra are plotted
in Figs. \ref{fig:optics-benzo}-\ref{fig:optics-naphtho-8}. Detailed
information related to the excited states contributing to the spectra
is presented in tables S1-S10 of the Supporting Information.

A careful examination of the spectra illustrates the following important
points: (a) spectra obtained from screened parameter are always red-shifted
as compared to the absorption spectra computed using standard parameters,
consistent with the similar shift observed for the HOMO-LUMO gap mentioned
above, (b) with the increasing size, the absorption spectrum are red-shifted
for the C\textsubscript{2v} symmetric coronene derivatives, and the
first peak of the optical spectra appears due to the absorption of
a photon, polarized along y-direction because of a transition from
their ground state (\textsuperscript{1}A\textsubscript{1}) to the
\textsuperscript{1}B\textsubscript{2} excited state characterized
predominantly by the singly excited configuration $\arrowvert H\rightarrow L\rangle$,
corresponding to the optical gap. For the C\textsubscript{s} symmetry
structure, the first peak representing the optical gap is due to the
absorption of a photon with mixed x-y polarization, to a state dominated
again by $\arrowvert H\rightarrow L\rangle$ configuration, (c) the
first peak obtained from the screened parameter calculations, is moderately
intense, whereas the standard parameter calculations predict the first
peak to be relatively of lower intensity for each molecule, (d) the
position of the first peak and higher energy peaks have significant
dependence on the Coulomb parameters employed in the calculations.

The experimental data for the optical gap is available only for benzo{[}a{]}coronene
and naphtho{[}8,1,2-abc{]}coronene molecules, presented in the last
column of the Table \ref{tab:optical-gap}. It is obvious from the
table that the PPP-CI approach predicts the optical gaps much more
accurately as compared to the TB model, and PPP-RHF approach. Therefore,
we only compare the PPP-CI results with the experimental data, wherever
available. The first peak (3.38 eV) of the experimentally obtained
spectra of benzo{[}a{]}coronene is in good agreement with the screened
parameter value (3.42 eV), as well as the standard parameter one (3.44
eV). For this case, TDDFT predicts a much lower value of 3.13 eV,
which clearly underestimates the experimental band gap by about 0.25
eV. For naphtho{[}8,1,2-abc{]}coronene also, the experimental value
of the optical gap (3.02 eV) is in good agreement with the screened
parameter value (2.95 eV), but standard parameter value (3.23 eV)
is significantly higher. Here, TDDFT predicted value of 2.92 eV is
in good agreement both with the experiment, and our screened parameter
based PPP-CI value. This suggests that our screened parameter PPP-CI
values of the optical gaps of the two other molecules, for which no
experimental results are available, are likely to be close to the
true values. Next, we discuss our CI results for the higher energy
regions of the absorption spectra of individual molecules.

\subsubsection{Benzo{[}ghi{]}perylene (C\protect\textsubscript{22}H\protect\textsubscript{12})}

\begin{figure}[h]
\begin{centering}
\includegraphics[scale=0.4]{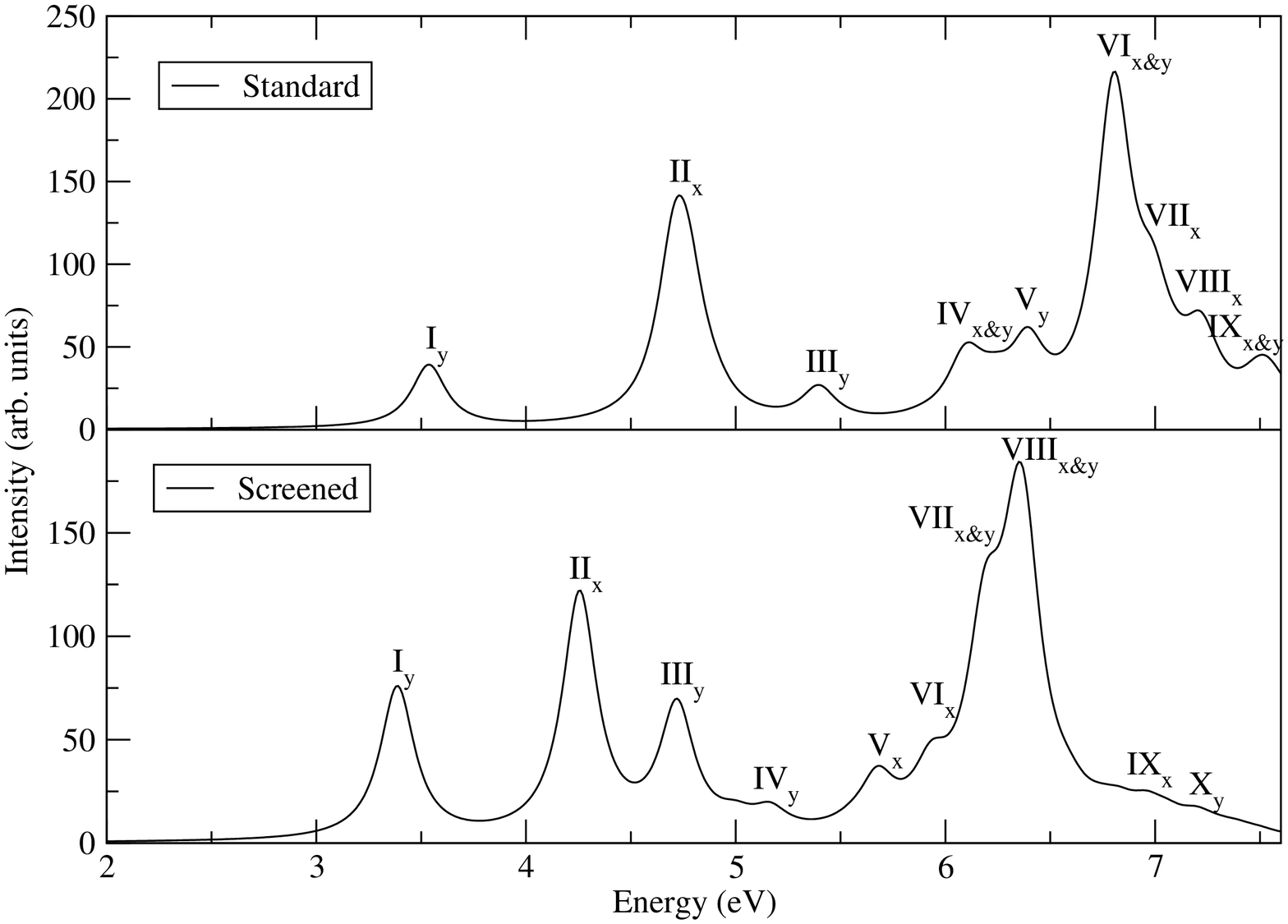}
\par\end{centering}
\caption{Linear optical absorption spectrum of benzo{[}ghi{]}perylene (C\protect\textsubscript{22}H\protect\textsubscript{12})
computed by employing PPP-QCI methodology. Both the Coulomb parameters,
screened, and standard, were separately used to compute the spectra.
Uniform line-width of 0.1 eV was adopted to plot the absorption spectra.
Subscripts of peak labels indicate the polarization direction of the
photon absorbed in the transition.\label{fig:optics-benzo-pery}}
\end{figure}

The geometry of benzo{[}ghi{]}perylene is presented in Fig. \ref{fig:geometry}(a),
whereas its optical absorption spectra are depicted in Fig. \ref{fig:optics-benzo-pery}.
We calculated the photoabsorption spectra by employing QCI methodology
and PPP model Hamiltonian employing both the screened, and the standard
Coulomb parameters. Detailed information related to the excited states
contributing to the peaks of the optical absorption spectra is presented
in tables S1 and S2 of the Supporting Information.

Fig. \ref{fig:optics-benzo-pery} reveals that the spectra, computed
using both the Coulomb parameters, have similar qualitative features.
Because, both the spectra begin with moderately intense peaks, and
the second peaks are intense peaks followed by several weaker peaks,
and the most intense peaks appear at the higher energy region. Experimental
data of the absorption spectra of this molecule is not available for
comparison.

Now, we discuss the two intense peaks of the optical spectra obtained
from both the Coulomb parameters. The second peak in both the spectra
is due to the excited states, whose wave functions are dominated by
the single excited configurations $\arrowvert H-1\rightarrow L\rangle$
and $\arrowvert H\rightarrow L+1\rangle$. The most intense peak located
at 6.35 eV (peak VIII, screened parameters) is due to two excited
states whose wave functions are largely composed of $\arrowvert H-2\rightarrow L+4\rangle$,
$\arrowvert H-4\rightarrow L+2\rangle$, $\arrowvert H-3\rightarrow L+4\rangle$,
and $\arrowvert H-4\rightarrow L+3\rangle$. The most intense peak
in the standard parameter spectrum near 6.8 eV (peak VI) is due to
three excited states whose wave functions are composed mainly of the
configurations $\arrowvert H-2\rightarrow L+3\rangle$, $\arrowvert H-3\rightarrow L+2\rangle$,
$\arrowvert H-3\rightarrow L+3\rangle$, $\arrowvert H-2\rightarrow L+2\rangle$,
$\arrowvert H-4\rightarrow L+2\rangle$, and $\arrowvert H-2\rightarrow L+4\rangle$.

\subsubsection{Benzo{[}a{]}coronene (C\protect\textsubscript{28}H\protect\textsubscript{14})}

\begin{figure}[H]
\begin{centering}
\includegraphics[scale=0.4]{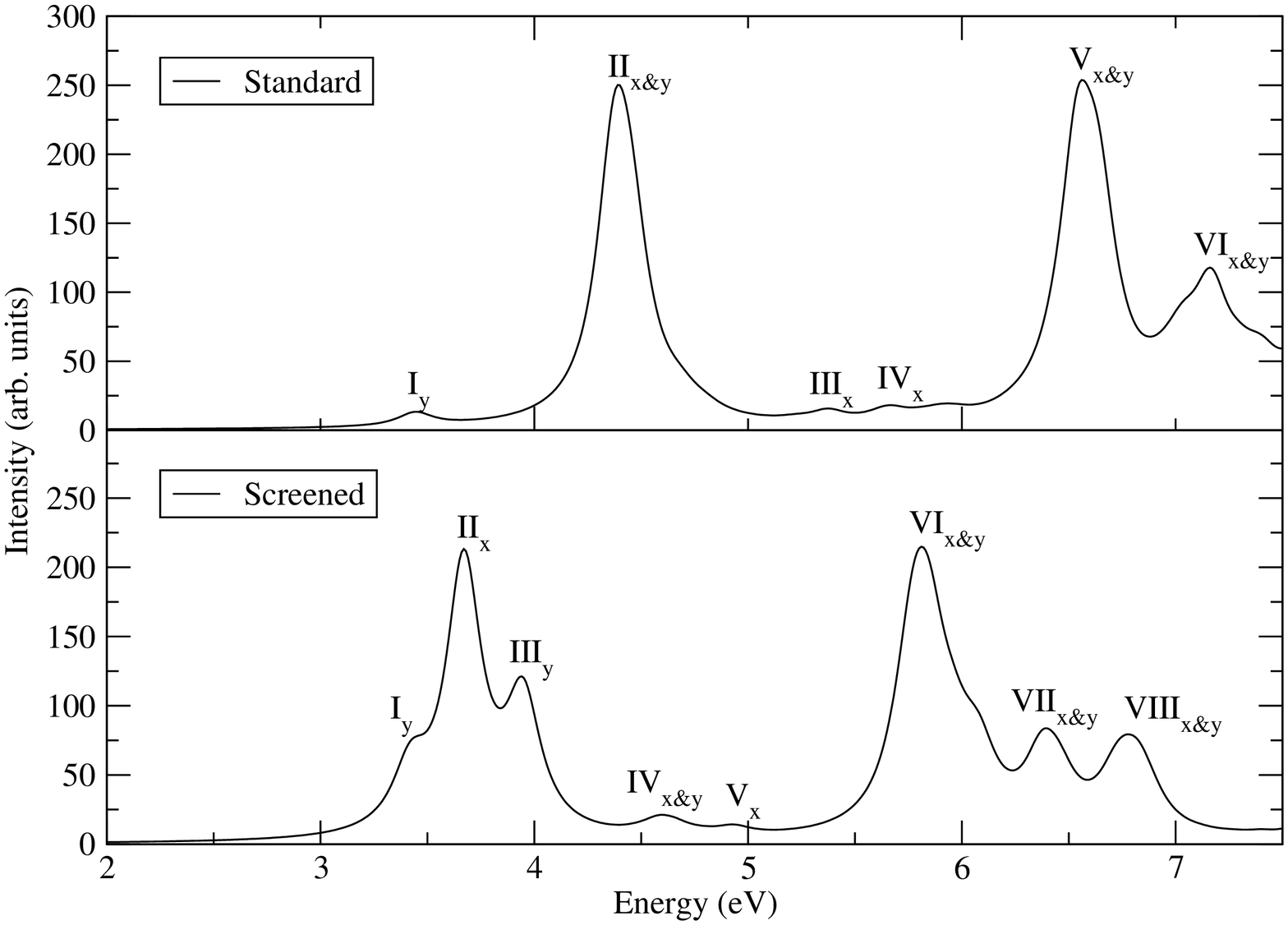}
\par\end{centering}
\caption{The linear optical absorption spectrum of benzo{[}a{]}coronene (C\protect\textsubscript{28}H\protect\textsubscript{14})
computed by employing our PPP-MRSDCI methodology. Both the Coulomb
parameters, screened and standard were separately used to compute
the spectra. Uniform line-width of 0.1 eV was adopted to plot the
absorption spectra. Subscripts of peak labels indicate the polarization
direction of the photon absorbed in the transition.\label{fig:optics-benzo}}
\end{figure}

The geometry of benzo{[}a{]}coronene is presented in Fig. \ref{fig:geometry}(b).
We computed the linear optical spectra of benzo{[}a{]}coronene, plotted
in Fig. \ref{fig:optics-benzo}, using both the Coulomb parameters,
screened and standard. The dominant configurations contributing to
the wave functions of the excited states of the corresponding peaks
are provided in tables S3-S4 of the Supporting Information.

From Fig. \ref{fig:optics-benzo}, it is obvious that, irrespective
of the Coulomb parameters, there are two major peaks in the spectra,
in addition to a number of less intense peaks. A solution-phase experimental
study of optical absorption in this molecule was performed by Bagley
\emph{et al}.\cite{Bagley_el_al} Our screened parameter peak positions
are compared to the measured ones in Table \ref{tab:expt-vs-theory},
and it is obvious that the two sets of values are in generally very
good agreement with each other. On the other hand, for these peaks,
the agreement between the experimental values and the standard parameter
results was poor. The optical absorption spectrum computed using the
standard parameters exhibits only one intense peak, in addition to
the first peak (peak I). While the location of peak I is in good agreement
with the experimental data as discussed earlier, but the disagreement
in the location of this second peak is close to 22\%. From Table \ref{tab:expt-vs-theory},
it is obvious that, except for two peaks located at 3.47 and 4.09
eV, the quantitative agreement between our screened parameter calculations
and the experiments is excellent. Our screened parameter calculations
also predict an intense peak (peak VI) near 5.80 eV; however, there
is no experimental data available beyond 5.39 eV. We hope that in
future experiments this higher energy region of the spectrum will
be probed, so as to verify whether or not the predictions of our computed
spectrum in that region hold.

Next, we analyze the wave functions of the excited states contributing
to the two most intense peaks in the computed spectra. Peak II of
both the spectra located at 3.67 eV (screened) and 4.38 eV (standard)
are dominated by the singly excited configurations $\arrowvert H\rightarrow L+1\rangle$,
and $\arrowvert H-1\rightarrow L\rangle$, while the second intense
features near 5.8 eV (peak VI, screened parameters) and 6.5 eV (peak
V of standard parameters) are due to states whose wave functions are
largely composed of the single excitations $\arrowvert H-3\rightarrow L+2\rangle$,
and $\arrowvert H-2\rightarrow L+3\rangle$. For the weaker peaks
of the absorption spectra also the contributions to the wave functions
of excited states are derived mainly from the singly excited configurations.

\begin{table}[H]
\caption{Comparison of experimental peak locations with those calculated using
the screened parameters, in the absorption spectrum of benzo{[}a{]}coronene,
excluding the optical gap. \label{tab:expt-vs-theory}}

\centering{}%
\begin{tabular}{cccccc}
\hline 
Molecule & symmetry & \hspace{0.3cm} & Peak position (eV) & \hspace{0.2cm} & Peak position (eV)\tabularnewline
 &  &  & Expt. work\cite{Bagley_el_al} &  & This work (Theory)\tabularnewline
\hline 
\hline 
Benzo{[}a{]}coronene & - &  & 3.47 &  & -\tabularnewline
(C\textsubscript{28}H\textsubscript{14}) & \textsuperscript{1}A\textsubscript{1} &  & 3.61 &  & 3.67\tabularnewline
 & \textsuperscript{1}B\textsubscript{2} &  & 3.95 &  & 3.95\tabularnewline
 & - &  & 4.09 &  & -\tabularnewline
 & \textsuperscript{1}B\textsubscript{2} &  & 4.51 &  & 4.58\tabularnewline
 & \textsuperscript{1}A\textsubscript{1} &  & 4.65 &  & 4.66\tabularnewline
\hline 
\end{tabular}
\end{table}

\subsubsection{Naphtho{[}2,3a{]}coronene (C\protect\textsubscript{32}H\protect\textsubscript{16})}

\begin{figure}[H]
\begin{centering}
\includegraphics[scale=0.4]{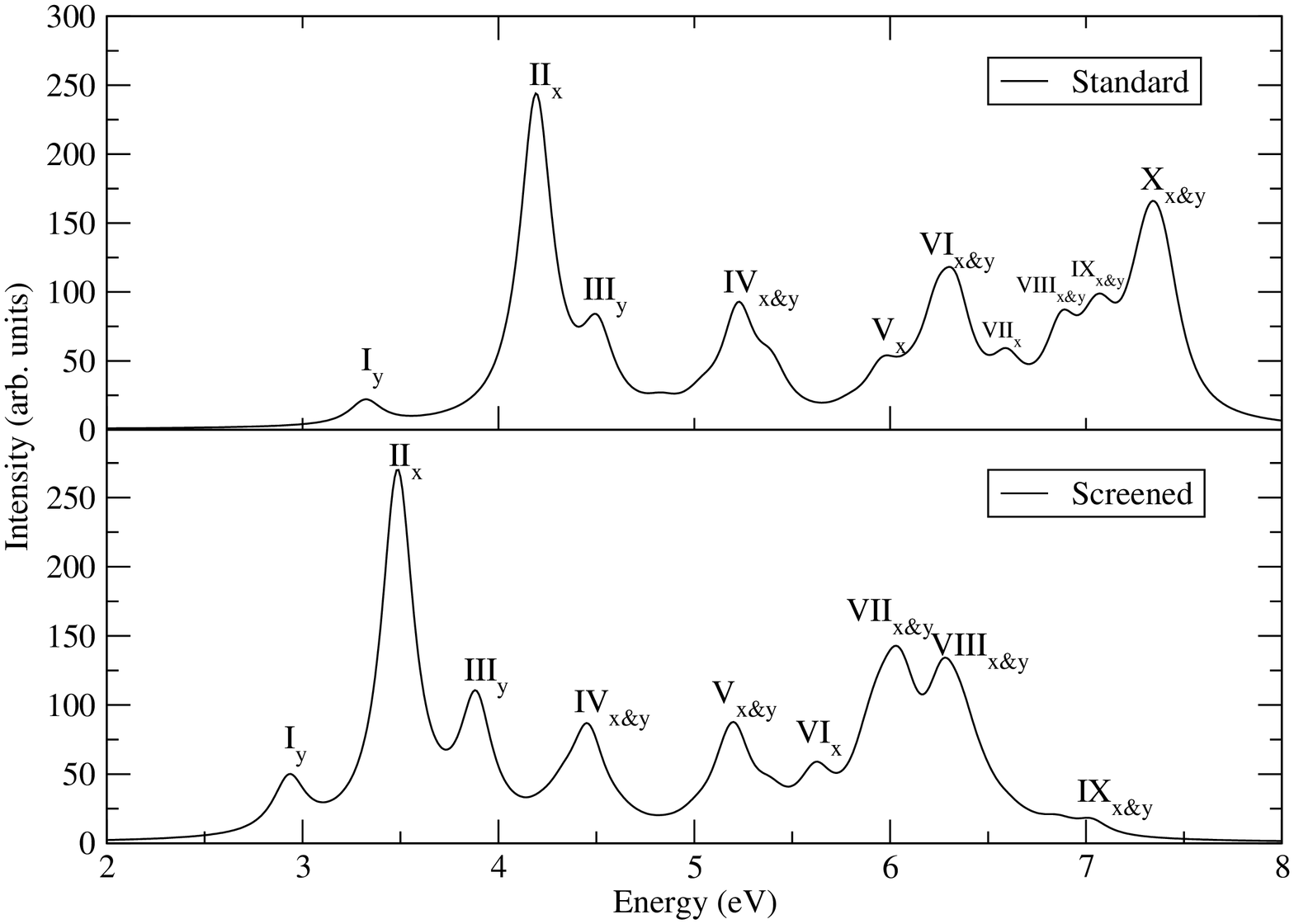}
\par\end{centering}
\caption{The linear optical absorption spectrum of naphtho{[}2,3a{]}coronene
(C\protect\textsubscript{32}H\protect\textsubscript{16}) computed
by employing our PPP-MRSDCI methodology. Both the Coulomb parameters,
screened and standard were separately used to compute the spectra.
Uniform line-width of 0.1 eV was adopted to plot the absorption spectra.
Subscripts of peak labels indicate the polarization direction of the
photon absorbed in the transition.\label{fig:optics-naphtho-2}}
\end{figure}

The geometry of the molecule is presented in Fig. \ref{fig:geometry}(c),
whereas the Fig. \ref{fig:optics-naphtho-2} represents the linear
optical photoabsorption spectra of naphtho{[}2,3a{]}coronene computed
using the PPP-CI approach with screened and standard parameters, separately.
The configurations with significant contribution to the peaks of the
optical spectra are presented in tables S5 and S6 of the Supporting
Information with quantitative descriptions of some other parameters.

The spectra obtained using both sets of Coulomb parameters start with
a very similar trend, and the second peaks of the absorption spectra,
which are the most intense peaks, are followed by several weaker peaks.
For this molecule, no experimental data is available for comparison
with our computed photoabsorption spectra.

Next, we discuss the wave functions of the excited states contributing
to intense peaks of the computed spectra. The most intense peak (peak-II)
of both the spectra near 3.48 eV (screened) and 4.19 eV (standard)
appear due to the states whose wave functions are characterized by
the two equally contributing singly-excited configurations $\arrowvert H\rightarrow L+1\rangle$
and $\arrowvert H-1\rightarrow L\rangle$. Peak III near 3.88 eV (screened)
and 4.50 eV (standard) are due to transitions to states whose wave
functions are dominated by the single excitation $\arrowvert H-1\rightarrow L+1\rangle$
with the photon polarized along the $y$-direction. In the screened
parameter spectrum, peaks VII (6.00 eV) and VIII (6.30 eV) are the
last two closely spaced intense peaks. For peak VII the excited state
wave function is dominated by single excitation $\arrowvert H-3\rightarrow L+3\rangle$
and double excitations $\arrowvert H-2\rightarrow L;H-1\rightarrow L\rangle$
and $\arrowvert H\rightarrow L+2;H\rightarrow L+1\rangle$. The excited
state corresponding to peak VIII, on the other hand, is strictly composed
of the single excitations $\arrowvert H-4\rightarrow L+5\rangle$
and $\arrowvert H-5\rightarrow L+4\rangle$. The last peak (peak X,
standard parameters) near 7.4 eV is due to a state whose wave functions
are mainly derived from the singly excited configurations $\arrowvert H-3\rightarrow L+6\rangle$,
$\arrowvert H-6\rightarrow L+3\rangle$, $\arrowvert H-9\rightarrow L+1\rangle$
and $\arrowvert H-1\rightarrow L+9\rangle$.

\subsubsection{Anthra{[}2,3a{]}coronene (C\protect\textsubscript{36}H\protect\textsubscript{18})}

\begin{figure}[H]
\begin{centering}
\includegraphics[scale=0.4]{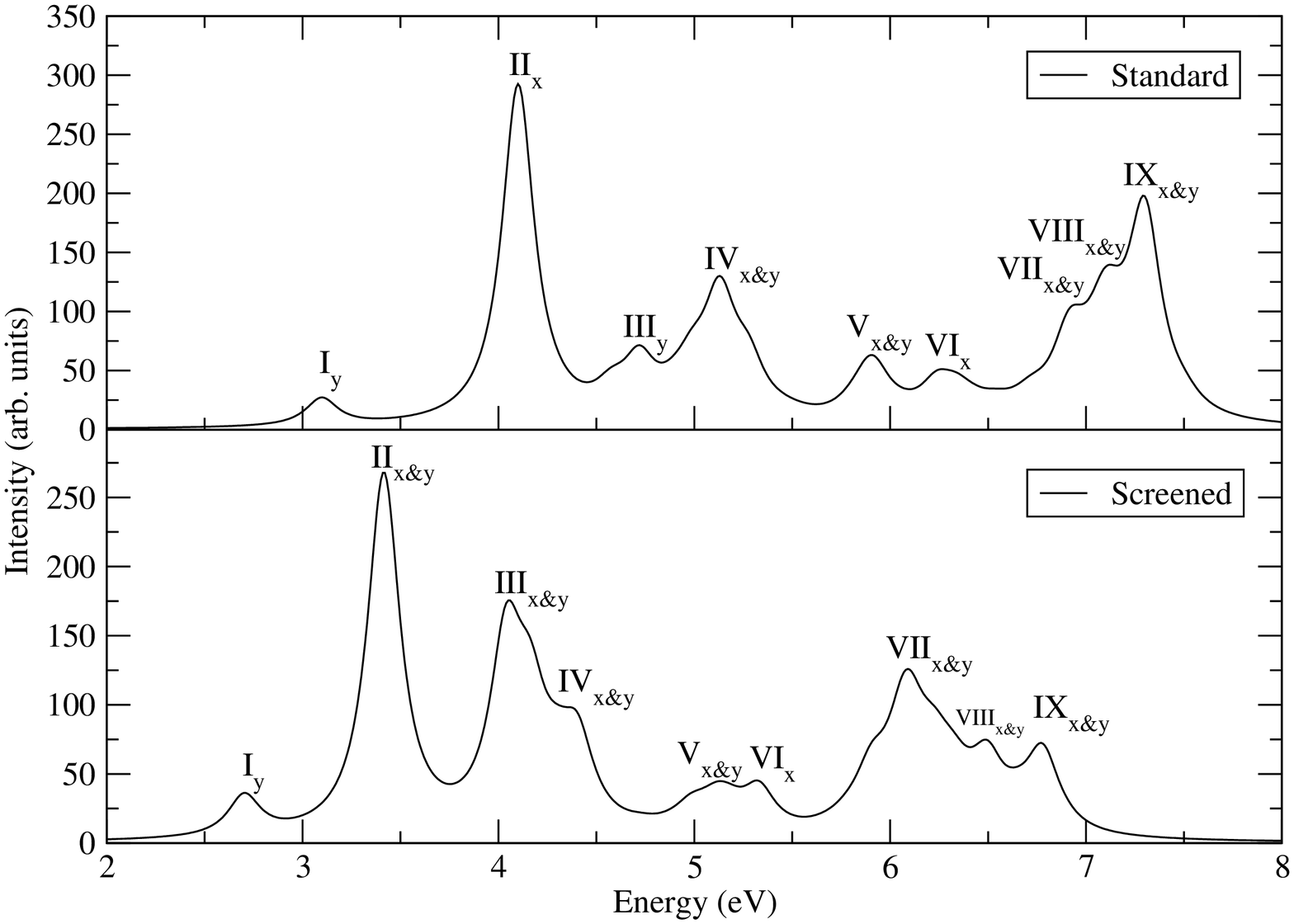}
\par\end{centering}
\caption{The linear optical absorption spectrum of anthra{[}2,3a{]}coronene
(C\protect\textsubscript{36}H\protect\textsubscript{18}) computed
by employing our PPP-MRSDCI methodology. Both the Coulomb parameters,
screened and standard were separately used to compute the spectra.
Uniform line-width of 0.1 eV was adopted to plot the absorption spectra.
Subscripts of peak labels indicate the polarization direction of the
photon absorbed in the transition.\label{fig:optics-anthra}}
\end{figure}

Fig. \ref{fig:optics-anthra} presents the optical spectra of anthra{[}2,3a{]}coronene
molecule calculated using PPP-CI methodology with both, screened and
standard parameters. The dominant configurations contributing to the
wave functions of the excited states of the corresponding peaks are
provided in tables S7 and S8 of the Supporting Information.

The spectra computed using the two sets of Coulomb parameters have
similar qualitative features in that both start with a small peak,
followed by the maximum intensity peak. The higher energy regions
of the two spectra contain a series of low to moderate intensity peaks.
For this molecule also no experimental data on its optical absorption
is available.

The many-body wave functions of the excited states corresponding to
the second peak of both the spectra, which are the most intense ones,
are dominated by the equally contributing single excitations $\arrowvert H\rightarrow L+2\rangle$,
and $\arrowvert H-2\rightarrow L\rangle$, with the polarization along
$x$-direction. Peaks III of both the spectra appear near 4.1 eV (screened)
and 4.7 eV (standard) due to the excited states whose many-body wave
functions are composed of equally contributing singly-excited configurations
$\arrowvert H\rightarrow L+4\rangle$, and $\arrowvert H-4\rightarrow L\rangle$.
Moderately intense peak IX at 7.3 eV (standard) is dominated by the
singly excited configurations $\arrowvert H-3\rightarrow L+3\rangle$
and $\arrowvert H-4\rightarrow L+4\rangle$, whereas peak VIII at
6.5 eV (screened) exhibits strong mixing of single and double excitations
$\arrowvert H-6\rightarrow L+6\rangle$, $\arrowvert H-5\rightarrow L+5\rangle$,
$\arrowvert H\rightarrow L+1;H\rightarrow L+3\rangle$, and $\arrowvert H-1\rightarrow L;H-3\rightarrow L\rangle$.

\subsubsection{Naphtho{[}8,1,2-abc{]}coronene (C\protect\textsubscript{30}H\protect\textsubscript{14})}

\begin{figure}[H]
\begin{centering}
\includegraphics[scale=0.4]{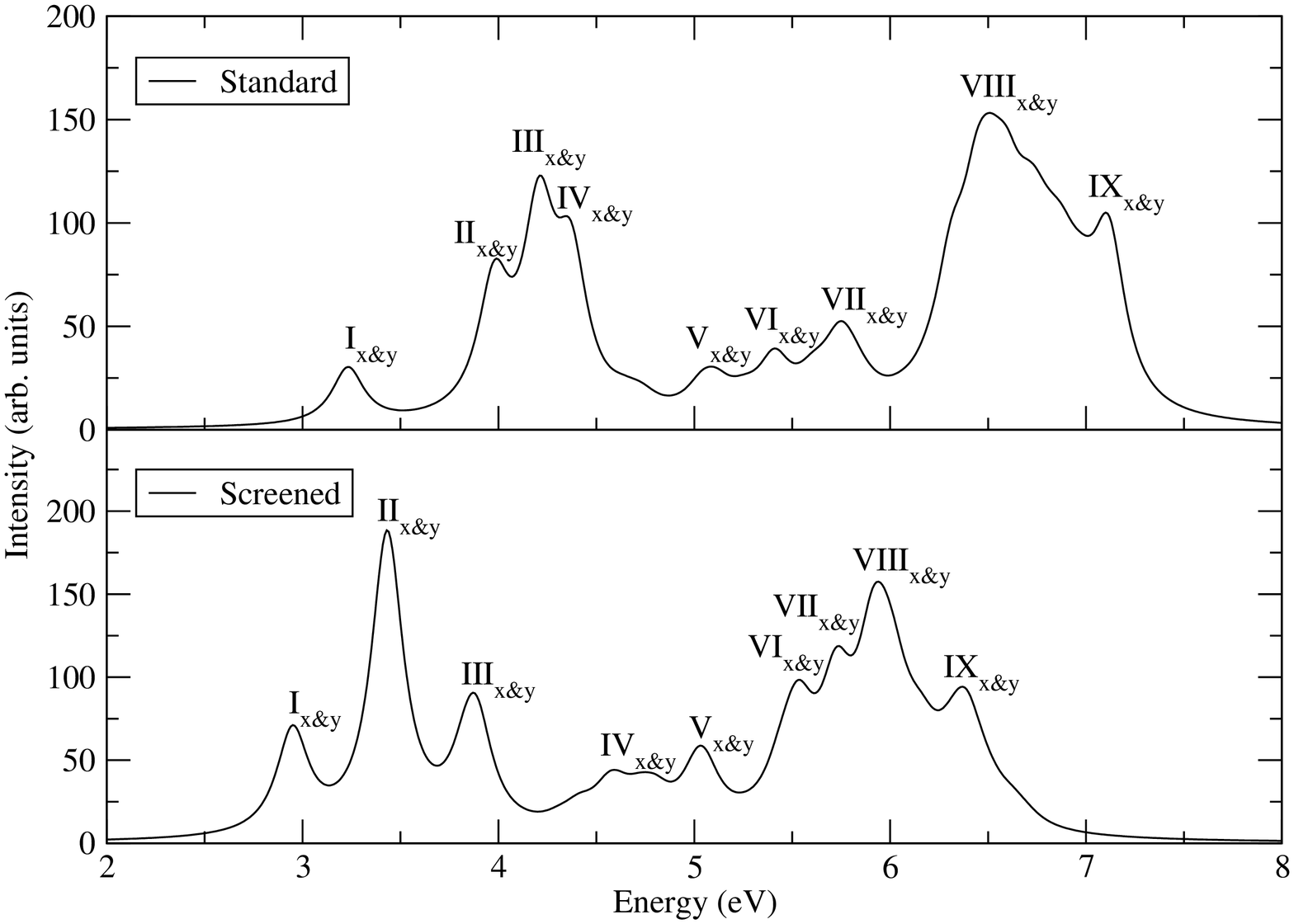}
\par\end{centering}
\caption{The linear optical absorption spectrum of naphtho{[}8,1,2-abc{]}coronene
(C\protect\textsubscript{30}H\protect\textsubscript{14}) computed
by employing our PPP-MRSDCI methodology. Both the Coulomb parameters,
screened and standard were separately used to compute the spectra.
Uniform line-width of 0.1 eV was adopted to plot the absorption spectra.
Subscripts of peak labels indicate the polarization direction of the
photon absorbed in the transition.\label{fig:optics-naphtho-8}}
\end{figure}

The linear optical spectra of naphtho{[}8,1,2-abc{]}coronene is presented
in Fig. \ref{fig:optics-naphtho-8} computed using both, screened,
and standard parameters within the framework of PPP-CI methodology.
The quantitative descriptions of the configurations of the wave functions
contributing to the peaks of the optical spectrum are presented in
tables S9 and S10 of the Supporting Information. The results of calculations
performed using the screened parameters are compared with the experimental
data in Table \ref{tab:expt-vs-theory-1}.

\begin{table}[H]
\caption{Comparison of experimental peak locations with those calculated using
the screened parameters, in the absorption spectrum of naphtho{[}8,1,2-abc{]}coronene,
excluding the optical gap. \label{tab:expt-vs-theory-1}}

\centering{}%
\begin{tabular}{cccccc}
\hline 
Molecule & symmetry & \hspace{0.3cm} & Peak position (eV) & \hspace{0.2cm} & Peak position (eV)\tabularnewline
 &  &  & Expt. work\cite{Bagley_el_al} &  & This work (Theory)\tabularnewline
\hline 
\hline 
Naphtho{[}8,1,2-abc{]}coronene & - &  & 3.19 &  & -\tabularnewline
(C\textsubscript{30}H\textsubscript{14}) & \textsuperscript{1}A &  & 3.34 &  & 3.43\tabularnewline
 & - &  & 3.62 &  & -\tabularnewline
 & - &  & 3.78 &  & -\tabularnewline
 & \textsuperscript{1}A &  & 3.95 &  & 3.89\tabularnewline
 & \textsuperscript{1}A &  & 4.75 - 4.83 &  & 4.56 - 4.72\tabularnewline
\hline 
\end{tabular}
\end{table}

From Fig. \ref{fig:optics-naphtho-8}, it is obvious that independent
of Coulomb parameters employed, both the spectra contain two intense
peaks, in addition to a number of weaker peaks. Bagley and Wornat
carried out an experimental study of optical absorption of this molecule
in the solution phase.\cite{Bagley_el_al} In Table \ref{tab:expt-vs-theory-1}
we have compared only the screened parameter based results with the
measured peaks,\cite{Bagley_el_al} because we obtain much better
agreement with those, as compared to standard parameter ones. For
the second peak of the experimental spectrum at 3.19 eV, we find no
candidate in our results. The location of the third experimental peak
(3.34 eV) is in decent agreement with the second peak (3.43 eV) of
our computed spectra. Then, two more experimental peaks (3.62 eV and
3.78 eV) are also absent in our calculations, but peak III of our
calculations at 3.89 eV is near the experimentally obtained peak near
3.95 eV. The calculated set of peaks in the range 4.56 - 4.72 eV slightly
underestimate the experimental peaks in the 4.75 - 4.83 eV energy
range.\cite{Bagley_el_al} Our calculated spectrum contains several
peaks in the higher energy region as well. However, the experimental
data terminates at 4.9 eV.\cite{Bagley_el_al} Therefore, we hope
that future experiments will extend to energies higher than this.

Now we discuss the wave functions of the excited states contributing
to the peaks of the absorption spectra. The most intense peaks near
3.4 eV (screened) and 4.2 eV (standard) of both the absorption spectra
are dominated by the singly excited configurations $\arrowvert H-1\rightarrow L\rangle$
and $\arrowvert H\rightarrow L+1\rangle$, but the intense peaks obtained
from the standard parameters are comparatively broader. The second
intense peaks of both the optical spectra near 5.9 eV (screened) and
6.6 eV (standard) are broader, and appear due to the states whose
wave functions consist mainly of single excitations $\arrowvert H-3\rightarrow L+4\rangle$,
$\arrowvert H-4\rightarrow L+3\rangle$, and $\arrowvert H-2\rightarrow L+6\rangle$,
$\arrowvert H-6\rightarrow L+2\rangle$, respectively.

\section{Conclusions}

\label{sec:conclusions}

We calculated the optical properties of five different PAH molecules,
namely, benzo{[}ghi{]}perylene (C\textsubscript{22}H\textsubscript{12}),
benzo{[}a{]}coronene (C\textsubscript{28}H\textsubscript{14}), naphtho{[}2,3a{]}coronene
(C\textsubscript{32}H\textsubscript{16}), anthra{[}2,3a{]}coronene
(C\textsubscript{36}H\textsubscript{18}), and naphtho{[}8,1,2-abc{]}coronene
(C\textsubscript{30}H\textsubscript{14}) by employing the large
scale electron-correlated PPP-CI methodology. For the sake of comparison,
we also computed the optical gaps of these molecules using first-principles
TDDFT. The common structural feature of these molecules is their similarity
to coronene. Our computed spectra of benzo{[}a{]}coronene and naphtho{[}8,1,2-abc{]}coronene
obtained by the PPP-CI approach are in good agreement with available
experimental data. We hope future experimental efforts to measure
the higher energy region of optical spectra of these two molecules,
and photoabsorption spectra of the other three molecules, \emph{i.e.},
benzo{[}ghi{]}perylene, naphtho{[}2,3a{]}coronene, anthra{[}2,3a{]}coronene,
against which our results could be benchmarked. A few important conclusions,
which can be drawn from our calculations, are:
\begin{enumerate}
\item Our screened parameter PPP-CI results predict moderately intense absorption
at the optical gap, while standard parameter predicts a much smaller
intensity.
\item For C\textsubscript{2v} symmetry coronene derivatives, the computed
optical spectra are red-shifted with the increasing size of the molecules,
and the distance between the first two peak locations are also gradually
increasing.
\item The computed optical spectra obtained using the screened parameters
are in better agreement with the experimental data as compared to
the standard parameter results.
\item The fact that we obtain good agreement of our PPP-CI calculations
with the available experimental data, implies that the $\sigma-\pi$
separation hypothesis which forms the basis of PPP model, is physically
correct when it comes to the description of optical transitions in
these molecules.
\item As compared to the coronene molecule, the optical gaps of all the
four coronene derivatives studied here are smaller.
\item Optical gaps computed by the TDDFT approach were found to be lower
than the corresponding PPP-CI values, for all the molecules.
\end{enumerate}
As far as future directions are concerned, it will be interesting
to probe photoinduced excited-state absorptions from the optical gap
of these molecules. Given the fact that none of these molecules have
inversion symmetry, it will be interesting to investigate whether
or not they exhibit second-order optical nonlinearities. Additionally,
electroabsorption spectra of these molecules may provide further useful
information about the nature of its low-lying excited states. Triplet
excited states of these molecules may shed some light on the influence
of electron correlation effects. At present, calculations along these
directions are underway in our group, and the results will be reported
in future works.

\section*{Supporting Information}

In the Supporting Information file, we first present a comparison
of the optical absorption spectra of $\hspace{1cm}$ benzo{[}ghi{]}perylene
molecule computed using the QCI and MRSDCI methods, employing the
screened parameters in the PPP model. This is followed by a series
of tables containing information about the excited states contributing
to important peaks in the calculated absorption spectra of the lower-symmetry
polycyclic aromatic hydrocarbon molecules considered in this work.

\section*{Author Information}

\subsection*{Corresponding Authors}

Alok Shukla:  {*}E-mail: shukla@phy.iitb.ac.in

\subsection*{Notes}

The authors declare no competing financial interests.

\section*{Acknowledgements}

Work of P.B. was supported by a Senior Research Fellowship offered
by University Grants Commission, India.

\section*{TOC Graphic}

\begin{figure}[H]
\includegraphics[scale=0.4]{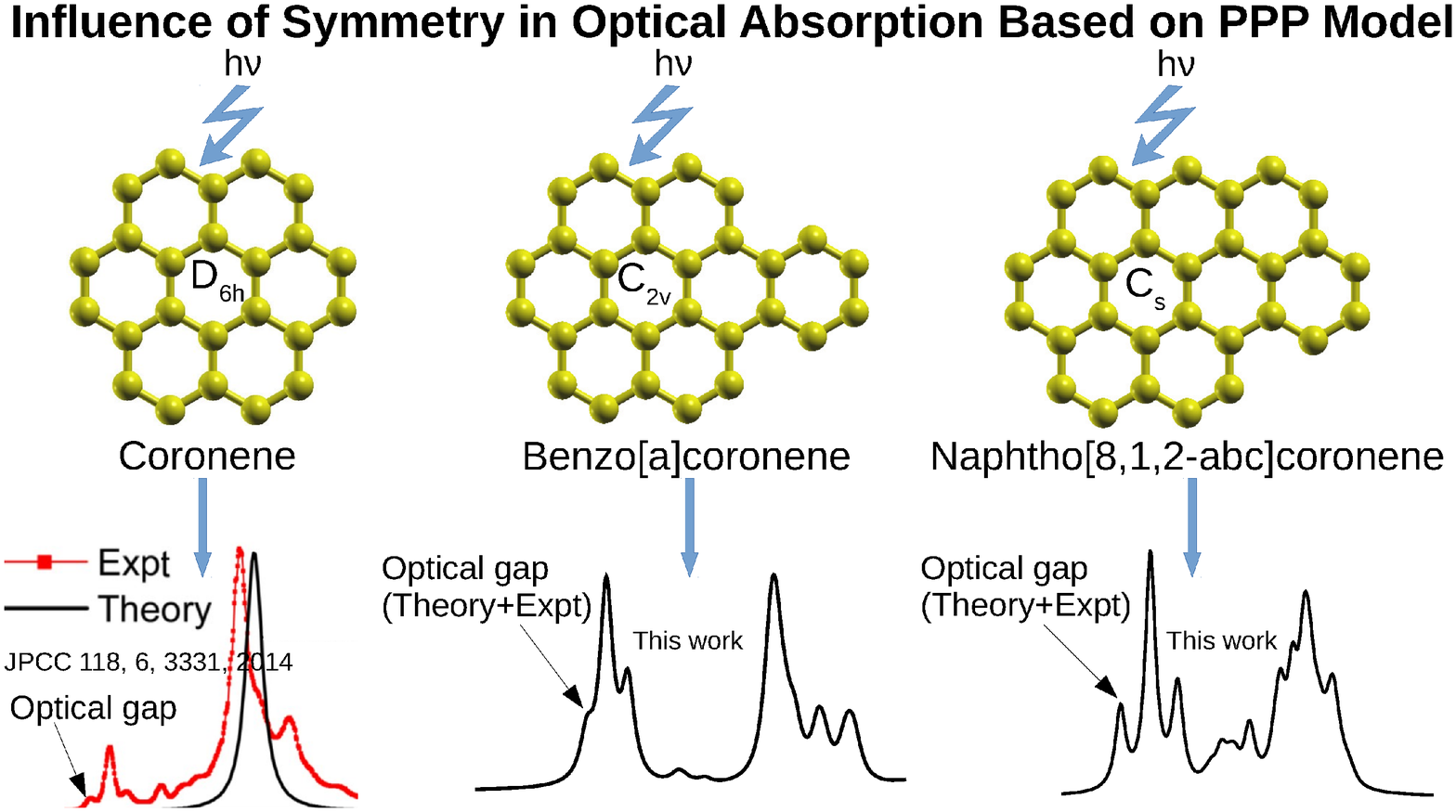}
\end{figure}

\bibliographystyle{achemso}
\addcontentsline{toc}{section}{\refname}\bibliography{coronene_group}

\end{document}


\title{Supporting Information:\\
Pariser-Parr-Pople Model based Configuration-Interaction Study of
Linear Optical Absorption in Lower-Symmetry Polycyclic Aromatic Hydrocarbon
Molecules}
\author{Pritam Bhattacharyya}
\email{pritambhattacharyya01@gmail.com}

\address{Department of Physics, Indian Institute of Technology Bombay, Powai,
Mumbai 400076, India}
\author{Deepak Kumar Rai}
\email{deepakrai@phy.iitb.ac.in}

\affiliation{Department of Physics, Indian Institute of Technology Bombay, Powai,
Mumbai 400076, India}
\affiliation{Present Address: Department of Chemistry, Southern University of Science
and Technology, Shenzhen, 518000, China}
\author{Alok Shukla}
\email{shukla@phy.iitb.ac.in}

\address{Department of Physics, Indian Institute of Technology Bombay, Powai,
Mumbai 400076, India}

\maketitle
In this supporting information, in Fig. S1, we first present a comparison
of the optical absorption spectra of benzo{[}ghi{]}perylene molecule
computed using the QCI and MRSDCI methods, employing the screened
parameters in the PPP model. This is followed by a series of tables
containing information about the excited states contributing to important
peaks in the calculated absorption spectra of the lower-symmetry polycyclic
aromatic hydrocarbon molecules considered in this work.

\begin{figure}[H]
\begin{centering}
\includegraphics[scale=0.45]{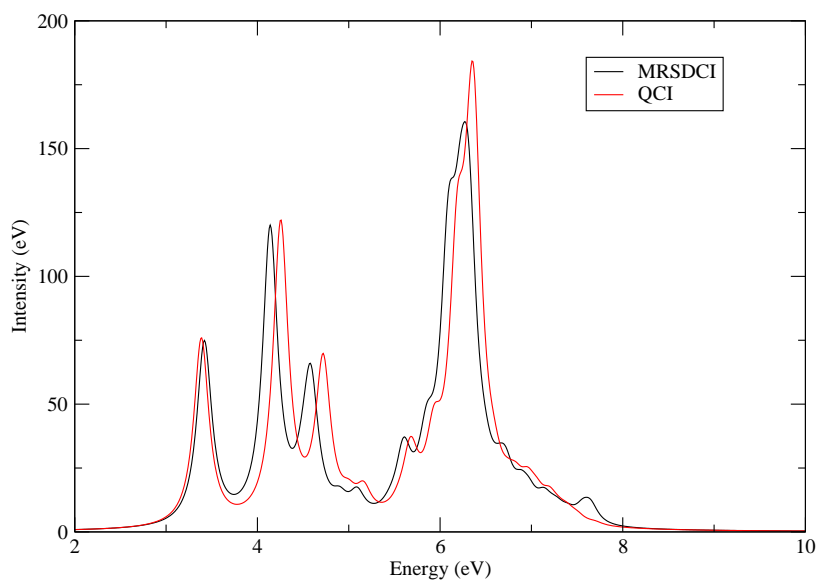}
\par\end{centering}
\caption{Optical absorption spectra of benzo{[}ghi{]}perylene molecule computed
using both the MRSDCI and QCI approaches, and the screened parameters,
employing the PPP model Hamiltonian. Uniform line width of 0.1 eV
was assumed while plotting both the spectra.}
\end{figure}

\begin{table}[H]
\caption{Many-particle wave functions of the excited states contributing to
the peaks in the linear optical absorption spectra of benzo{[}ghi{]}perylene
(C\protect\textsubscript{2v}) {[}Fig. 1a, main article{]} computed
using the screened parameters in the PPP semi-empirical method and
QCI methodology. In the 'Peak' column, $x$ ($y$) indicates that
the polarization of the absorbed photon is along $x$ ($y$) direction,
while $xy$ denotes mixed polarization.  The excited state symmetries
of the corresponding peaks are also presented below. $f$ (= $\frac{2\times m_{e}}{3\times\hbar^{2}}(E)\sum_{j}|\langle e|O_{j}|R\rangle|^{2}$)
indicate the oscillator strength for a particular transition where
$E$, $|e\rangle$, $|R\rangle$ and $O_{j}$ representing the peak
energy, excited states of the corresponding peak, reference state
and electric dipole operator for different Cartesian components, respectively.
In the last column, the fractional numbers inside the first bracket
represent the contributing coefficients to the CI wave functions for
each corresponding configuration.}

\begin{tabular}{ccccccccc}
\hline 
Peak & \hspace{1.7cm} & Symmetry & \hspace{1.7cm} & E (eV) & \hspace{1.7cm} & $f$ & \hspace{1.7cm} & Wave function\tabularnewline
\hline 
\hline 
 &  &  &  &  &  &  &  & \tabularnewline
I\textsubscript{$y$} &  & \textsuperscript{1}B\textsubscript{2} &  & 3.39 &  & 4.917 &  & $\arrowvert H\rightarrow L\rangle(0.8505)$\tabularnewline
 &  &  &  &  &  &  &  & {\scriptsize{}$\arrowvert H\rightarrow L;H-5\rightarrow L+1;H-1\rightarrow L+5\rangle(0.0684)$}\tabularnewline
 &  &  &  &  &  &  &  & \tabularnewline
II\textsubscript{$x$} &  & \textsuperscript{1}A\textsubscript{1} &  & 4.25 &  & 7.817 &  & $\arrowvert H-1\rightarrow L\rangle(0.5932)$\tabularnewline
 &  &  &  &  &  &  &  & $\arrowvert H\rightarrow L+1\rangle(0.5932)$\tabularnewline
 &  &  &  &  &  &  &  & \tabularnewline
III\textsubscript{$y$} &  & \textsuperscript{1}B\textsubscript{2} &  & 4.72 &  & 4.099 &  & $\arrowvert H-1\rightarrow L+1\rangle(0.8000)$\tabularnewline
 &  &  &  &  &  &  &  & $\arrowvert H\rightarrow L+2\rangle(0.1805)$\tabularnewline
 &  &  &  &  &  &  &  & \tabularnewline
IV\textsubscript{$y$} &  & \textsuperscript{1}B\textsubscript{2} &  & 5.16 &  & 0.667 &  & $\arrowvert H\rightarrow L+4\rangle(0.5369)$\tabularnewline
 &  &  &  &  &  &  &  & $\arrowvert H-4\rightarrow L\rangle(0.5369)$\tabularnewline
 &  &  &  &  &  &  &  & \tabularnewline
V\textsubscript{$x$} &  & \textsuperscript{1}A\textsubscript{1} &  & 5.67 &  & 1.627 &  & $\arrowvert H-4\rightarrow L+1\rangle(0.3766)$\tabularnewline
 &  &  &  &  &  &  &  & $\arrowvert H-1\rightarrow L+4\rangle(0.3766)$\tabularnewline
 &  &  &  &  &  &  &  & \tabularnewline
VI\textsubscript{$x$} &  & \textsuperscript{1}A\textsubscript{1} &  & 5.93 &  & 1.617 &  & $\arrowvert H-4\rightarrow L+1\rangle(0.3911)$\tabularnewline
 &  &  &  &  &  &  &  & $\arrowvert H-1\rightarrow L+4\rangle(0.3911)$\tabularnewline
 &  &  &  &  &  &  &  & \tabularnewline
VII\textsubscript{$xy$} &  & \textsuperscript{1}A\textsubscript{1} &  & 6.18 &  & 1.571 &  & $\arrowvert H-6\rightarrow L\rangle(0.3569)$\tabularnewline
 &  &  &  &  &  &  &  & $\arrowvert H\rightarrow L+6\rangle(0.3569)$\tabularnewline
 &  & \textsuperscript{1}B\textsubscript{2} &  & 6.19 &  & 4.080 &  & $\arrowvert H-3\rightarrow L+3\rangle(0.4944)$\tabularnewline
 &  &  &  &  &  &  &  & $\arrowvert H-2\rightarrow L+2\rangle(0.4592)$\tabularnewline
 &  &  &  &  &  &  &  & \tabularnewline
VIII\textsubscript{$xy$} &  & \textsuperscript{1}B\textsubscript{2} &  & 6.34 &  & 3.865 &  & $\arrowvert H-2\rightarrow L+4\rangle(0.4813)$\tabularnewline
 &  &  &  &  &  &  &  & $\arrowvert H-4\rightarrow L+2\rangle(0.4813)$\tabularnewline
 &  & \textsuperscript{1}A\textsubscript{1} &  & 6.37 &  & 6.752 &  & $\arrowvert H-3\rightarrow L+4\rangle(0.4718)$\tabularnewline
 &  &  &  &  &  &  &  & $\arrowvert H-4\rightarrow L+3\rangle(0.4718)$\tabularnewline
 &  &  &  &  &  &  &  & \tabularnewline
IX\textsubscript{$x$} &  & \textsuperscript{1}A\textsubscript{1} &  & 6.96 &  & 0.640 &  & $\arrowvert H-7\rightarrow L+1\rangle(0.4566)$\tabularnewline
 &  &  &  &  &  &  &  & $\arrowvert H-1\rightarrow L+7\rangle(0.4566)$\tabularnewline
 &  &  &  &  &  &  &  & \tabularnewline
X\textsubscript{$y$} &  & \textsuperscript{1}B\textsubscript{2} &  & 7.20 &  & 0.459 &  & $\arrowvert H-1\rightarrow L;H-1\rightarrow L+1\rangle(0.3154)$\tabularnewline
 &  &  &  &  &  &  &  & $\arrowvert H\rightarrow L+1;H-1\rightarrow L+1\rangle(0.3154)$\tabularnewline
\hline 
\end{tabular}
\end{table}

\begin{table}[H]
\caption{Many-particle wave functions of the excited states contributing to
the peaks in the linear optical absorption spectra of benzo{[}ghi{]}perylene
(C\protect\textsubscript{2v}) {[}Fig. 1a, main article{]} computed
using the standard parameters in the PPP semi-empirical method and
QCI methodology. In the 'Peak' column, $x$ ($y$) indicates that
the polarization of the absorbed photon is along $x$ ($y$) direction,
while $xy$ denotes mixed polarization. The excited state symmetries
of the corresponding peaks are also presented below. $f$ (= $\frac{2\times m_{e}}{3\times\hbar^{2}}(E)\sum_{j}|\langle e|O_{j}|R\rangle|^{2}$)
indicate the oscillator strength for a particular transition where
$E$, $|e\rangle$, $|R\rangle$ and $O_{j}$ representing the peak
energy, excited states of the corresponding peak, reference state
and electric dipole operator for different Cartesian components, respectively.
In the last column, the fractional numbers inside the first bracket
represent the contributing coefficients to the CI wave functions for
each corresponding configuration.}

\begin{tabular}{ccccccccc}
\hline 
Peak & \hspace{1.7cm} & Symmetry & \hspace{1.7cm} & E (eV) & \hspace{1.7cm} & $f$ & \hspace{1.7cm} & Wave function\tabularnewline
\hline 
\hline 
 &  &  &  &  &  &  &  & \tabularnewline
I\textsubscript{$y$} &  & \textsuperscript{1}B\textsubscript{2} &  & 3.54 &  & 2.519 &  & $\arrowvert H\rightarrow L\rangle(0.8354)$\tabularnewline
 &  &  &  &  &  &  &  & $\arrowvert H-1\rightarrow L+1\rangle(0.2493)$\tabularnewline
 &  &  &  &  &  &  &  & \tabularnewline
II\textsubscript{$x$} &  & \textsuperscript{1}A\textsubscript{1} &  & 4.71 &  & 5.994 &  & $\arrowvert H-1\rightarrow L\rangle(0.5449)$\tabularnewline
 &  &  &  &  &  &  &  & $\arrowvert H\rightarrow L+1\rangle(0.5449)$\tabularnewline
 &  &  &  &  &  &  &  & \tabularnewline
III\textsubscript{$y$} &  & \textsuperscript{1}B\textsubscript{2} &  & 5.40 &  & 1.361 &  & $\arrowvert H\rightarrow L+4\rangle(0.4600)$\tabularnewline
 &  &  &  &  &  &  &  & $\arrowvert H-4\rightarrow L\rangle(0.4600)$\tabularnewline
 &  &  &  &  &  &  &  & \tabularnewline
IV\textsubscript{$xy$} &  & \textsuperscript{1}A\textsubscript{1} &  & 6.09 &  & 1.847 &  & $\arrowvert H-4\rightarrow L+1\rangle(0.4128)$\tabularnewline
 &  &  &  &  &  &  &  & $\arrowvert H-1\rightarrow L+4\rangle(0.4128)$\tabularnewline
 &  & \textsuperscript{1}B\textsubscript{2} &  & 6.13 &  & 0.698 &  & $\arrowvert H-2\rightarrow L+2\rangle(0.4852)$\tabularnewline
 &  &  &  &  &  &  &  & $\arrowvert H-3\rightarrow L+3\rangle(0.4465)$\tabularnewline
 &  &  &  &  &  &  &  & \tabularnewline
V\textsubscript{$y$} &  & \textsuperscript{1}B\textsubscript{2} &  & 6.39 &  & 2.586 &  & $\arrowvert H\rightarrow L+7\rangle(0.3035)$\tabularnewline
 &  &  &  &  &  &  &  & $\arrowvert H-7\rightarrow L\rangle(0.3035)$\tabularnewline
 &  &  &  &  &  &  &  & \tabularnewline
VI\textsubscript{$xy$} &  & \textsuperscript{1}A\textsubscript{1} &  & 6.79 &  & 6.480 &  & $\arrowvert H-2\rightarrow L+3\rangle(0.3681)$\tabularnewline
 &  &  &  &  &  &  &  & $\arrowvert H-3\rightarrow L+2\rangle(0.3681)$\tabularnewline
 &  & \textsuperscript{1}B\textsubscript{2} &  & 6.79 &  & 1.798 &  & $\arrowvert H-3\rightarrow L+3\rangle(0.4971)$\tabularnewline
 &  &  &  &  &  &  &  & $\arrowvert H-2\rightarrow L+2\rangle(0.3856)$\tabularnewline
 &  & \textsuperscript{1}B\textsubscript{2} &  & 6.82 &  & 5.024 &  & $\arrowvert H-4\rightarrow L+2\rangle(0.4389)$\tabularnewline
 &  &  &  &  &  &  &  & $\arrowvert H-2\rightarrow L+4\rangle(0.4389)$\tabularnewline
 &  &  &  &  &  &  &  & \tabularnewline
VII\textsubscript{$x$} &  & \textsuperscript{1}A\textsubscript{1} &  & 6.99 &  & 3.588 &  & $\arrowvert H-3\rightarrow L+4\rangle(0.5125)$\tabularnewline
 &  &  &  &  &  &  &  & $\arrowvert H-4\rightarrow L+3\rangle(0.5125)$\tabularnewline
 &  &  &  &  &  &  &  & \tabularnewline
VIII\textsubscript{$x$} &  & \textsuperscript{1}A\textsubscript{1} &  & 7.22 &  & 2.766 &  & $\arrowvert H\rightarrow L;H\rightarrow L+2\rangle(0.3294)$\tabularnewline
 &  &  &  &  &  &  &  & $\arrowvert H\rightarrow L;H-2\rightarrow L\rangle(0.3294)$\tabularnewline
 &  &  &  &  &  &  &  & \tabularnewline
IX\textsubscript{$x$} &  & \textsuperscript{1}A\textsubscript{1} &  & 7.52 &  & 0.709 &  & $\arrowvert H-7\rightarrow L+1\rangle(0.3811)$\tabularnewline
 &  &  &  &  &  &  &  & $\arrowvert H-1\rightarrow L+7\rangle(0.3811)$\tabularnewline
 &  & \textsuperscript{1}B\textsubscript{2} &  & 7.52 &  & 0.463 &  & $\arrowvert H-5\rightarrow L+3\rangle(0.2476)$\tabularnewline
 &  &  &  &  &  &  &  & $\arrowvert H-3\rightarrow L+5\rangle(0.2476)$\tabularnewline
 &  & \textsuperscript{1}B\textsubscript{2} &  & 7.56 &  & 0.576 &  & $\arrowvert H-1\rightarrow L;H-1\rightarrow L+1\rangle(0.3021)$\tabularnewline
 &  &  &  &  &  &  &  & $\arrowvert H\rightarrow L+1;H-1\rightarrow L+1\rangle(0.3021)$\tabularnewline
\hline 
\end{tabular}
\end{table}

\begin{table}[H]
\caption{Many-particle wave functions of the excited states contributing to
the peaks in the linear optical absorption spectra of benzo{[}a{]}coronene
(C\protect\textsubscript{2v}) {[}Fig. 1b, main article{]} computed
using the screened parameters in the PPP semi-empirical method and
MRSDCI methodology. In the 'Peak' column, $x$ ($y$) indicates that
the polarization of the absorbed photon is along $x$ ($y$) direction,
while $xy$ denotes mixed polarization. The excited state symmetries
of the corresponding peaks are also presented below. $f$ (= $\frac{2\times m_{e}}{3\times\hbar^{2}}(E)\sum_{j}|\langle e|O_{j}|R\rangle|^{2}$)
indicate the oscillator strength for a particular transition where
$E$, $|e\rangle$, $|R\rangle$ and $O_{j}$ representing the peak
energy, excited states of the corresponding peak, reference state
and electric dipole operator for different Cartesian components, respectively.
In the last column, the fractional numbers inside the first bracket
represent the contributing coefficients to the CI wave functions for
each corresponding configuration.}

\begin{tabular}{ccccccccc}
\hline 
Peak & \hspace{1.7cm} & Symmetry & \hspace{1.7cm} & E (eV) & \hspace{1.7cm} & $f$ & \hspace{1.7cm} & Wave function\tabularnewline
\hline 
\hline 
 &  &  &  &  &  &  &  & \tabularnewline
I\textsubscript{$y$} &  & \textsuperscript{1}B\textsubscript{2} &  & 3.42 &  & 2.898 &  & $\arrowvert H\rightarrow L\rangle(0.8658)$\tabularnewline
 &  &  &  &  &  &  &  & $\arrowvert H-1\rightarrow L+1\rangle(0.1582)$\tabularnewline
 &  &  &  &  &  &  &  & \tabularnewline
II\textsubscript{$x$} &  & \textsuperscript{1}A\textsubscript{1} &  & 3.67 &  & 13.010 &  & $\arrowvert H\rightarrow L+1\rangle(0.6075)$\tabularnewline
 &  &  &  &  &  &  &  & $\arrowvert H-1\rightarrow L\rangle(0.6075)$\tabularnewline
 &  &  &  &  &  &  &  & \tabularnewline
III\textsubscript{$y$} &  & \textsuperscript{1}B\textsubscript{2} &  & 3.95 &  & 6.345 &  & $\arrowvert H-1\rightarrow L+1\rangle(0.8540)$\tabularnewline
 &  &  &  &  &  &  &  & $\arrowvert H\rightarrow L\rangle(0.1473)$\tabularnewline
 &  &  &  &  &  &  &  & \tabularnewline
IV\textsubscript{$xy$} &  & \textsuperscript{1}B\textsubscript{2} &  & 4.58 &  & 0.608 &  & $\arrowvert H\rightarrow L+2\rangle(0.5913)$\tabularnewline
 &  &  &  &  &  &  &  & $\arrowvert H-2\rightarrow L\rangle(0.5913)$\tabularnewline
 &  & \textsuperscript{1}A\textsubscript{1} &  & 4.66 &  & 0.362 &  & $\arrowvert H-3\rightarrow L\rangle(0.5379)$\tabularnewline
 &  &  &  &  &  &  &  & $\arrowvert H\rightarrow L+3\rangle(0.5379)$\tabularnewline
 &  &  &  &  &  &  &  & \tabularnewline
V\textsubscript{$x$} &  & \textsuperscript{1}A\textsubscript{1} &  & 4.93 &  & 0.385 &  & $\arrowvert H-1\rightarrow L+2\rangle(0.4674)$\tabularnewline
 &  &  &  &  &  &  &  & $\arrowvert H-2\rightarrow L+1\rangle(0.4674)$\tabularnewline
 &  &  &  &  &  &  &  & \tabularnewline
VI\textsubscript{$xy$} &  & \textsuperscript{1}A\textsubscript{1} &  & 5.78 &  & 6.957 &  & $\arrowvert H-3\rightarrow L+2\rangle(0.4952)$\tabularnewline
 &  &  &  &  &  &  &  & $\arrowvert H-2\rightarrow L+3\rangle(0.4952)$\tabularnewline
 &  & \textsuperscript{1}B\textsubscript{2} &  & 5.83 &  & 2.919 &  & $\arrowvert H-4\rightarrow L+2\rangle(0.5333)$\tabularnewline
 &  &  &  &  &  &  &  & $\arrowvert H-2\rightarrow L+4\rangle(0.5333)$\tabularnewline
 &  & \textsuperscript{1}B\textsubscript{2} &  & 6.08 &  & 2.726 &  & $\arrowvert H-3\rightarrow L+3\rangle(0.4925)$\tabularnewline
 &  &  &  &  &  &  &  & $\arrowvert H-1\rightarrow L+7\rangle(0.3147)$\tabularnewline
 &  &  &  &  &  &  &  & \tabularnewline
VII\textsubscript{$xy$} &  & \textsuperscript{1}B\textsubscript{2} &  & 6.37 &  & 2.983 &  & $\arrowvert H-1\rightarrow L+7\rangle(0.4461)$\tabularnewline
 &  &  &  &  &  &  &  & $\arrowvert H-7\rightarrow L+1\rangle(0.4461)$\tabularnewline
 &  & \textsuperscript{1}A\textsubscript{1} &  & 6.45 &  & 1.825 &  & $\arrowvert H-8\rightarrow L\rangle(0.3886)$\tabularnewline
 &  &  &  &  &  &  &  & $\arrowvert H\rightarrow L+8\rangle(0.3886)$\tabularnewline
 &  &  &  &  &  &  &  & \tabularnewline
VIII\textsubscript{$xy$} &  & \textsuperscript{1}A\textsubscript{1} &  & 6.71 &  & 0.727 &  & $\arrowvert H-3\rightarrow L+6\rangle(0.5157)$\tabularnewline
 &  &  &  &  &  &  &  & $\arrowvert H-6\rightarrow L+3\rangle(0.5157)$\tabularnewline
 &  & \textsuperscript{1}B\textsubscript{2} &  & 6.83 &  & 2.254 &  & $\arrowvert H-2\rightarrow L+6\rangle(0.3769)$\tabularnewline
 &  &  &  &  &  &  &  & $\arrowvert H-6\rightarrow L+2\rangle(0.3769)$\tabularnewline
\hline 
\end{tabular}
\end{table}

\begin{table}[H]
\caption{Many-particle wave functions of the excited states contributing to
the peaks in the linear optical absorption spectra of benzo{[}a{]}coronene
(C\protect\textsubscript{2v}) {[}Fig. 1b, main article{]} computed
using the standard parameters in the PPP semi-empirical method and
MRSDCI methodology. In the 'Peak' column, $x$ ($y$) indicates that
the polarization of the absorbed photon is along $x$ ($y$) direction,
while $xy$ denotes mixed polarization. The excited state symmetries
of the corresponding peaks are also presented below. $f$ (= $\frac{2\times m_{e}}{3\times\hbar^{2}}(E)\sum_{j}|\langle e|O_{j}|R\rangle|^{2}$)
indicate the oscillator strength for a particular transition where
$E$, $|e\rangle$, $|R\rangle$ and $O_{j}$ representing the peak
energy, excited states of the corresponding peak, reference state
and electric dipole operator for different Cartesian components, respectively.
In the last column, the fractional numbers inside the first bracket
represent the contributing coefficients to the CI wave functions for
each corresponding configuration.}

\begin{tabular}{ccccccccc}
\hline 
Peak & \hspace{1.7cm} & Symmetry & \hspace{1.7cm} & E (eV) & \hspace{1.7cm} & $f$ & \hspace{1.7cm} & Wave function\tabularnewline
\hline 
\hline 
 &  &  &  &  &  &  &  & \tabularnewline
I\textsubscript{$y$} &  & \textsuperscript{1}B\textsubscript{2} &  & 3.44 &  & 0.649 &  & $\arrowvert H\rightarrow L\rangle(0.7528)$\tabularnewline
 &  &  &  &  &  &  &  & $\arrowvert H-1\rightarrow L+1\rangle(0.4314)$\tabularnewline
 &  &  &  &  &  &  &  & \tabularnewline
II\textsubscript{$xy$} &  & \textsuperscript{1}A\textsubscript{1} &  & 4.38 &  & 12.181 &  & $\arrowvert H\rightarrow L+1\rangle(0.5829)$\tabularnewline
 &  &  &  &  &  &  &  & $\arrowvert H-1\rightarrow L\rangle(0.5829)$\tabularnewline
 &  & \textsuperscript{1}B\textsubscript{2} &  & 4.45 &  & 6.315 &  & $\arrowvert H-1\rightarrow L+1\rangle(0.6870)$\tabularnewline
 &  &  &  &  &  &  &  & $\arrowvert H\rightarrow L\rangle(0.3860)$\tabularnewline
 &  &  &  &  &  &  &  & \tabularnewline
III\textsubscript{$x$} &  & \textsuperscript{1}A\textsubscript{1} &  & 5.37 &  & 0.508 &  & $\arrowvert H-3\rightarrow L\rangle(0.3829)$\tabularnewline
 &  &  &  &  &  &  &  & $\arrowvert H\rightarrow L+3\rangle(0.3829)$\tabularnewline
 &  &  &  &  &  &  &  & \tabularnewline
IV\textsubscript{$x$} &  & \textsuperscript{1}A\textsubscript{1} &  & 5.67 &  & 0.442 &  & $\arrowvert H-1\rightarrow L+2\rangle(0.4246)$\tabularnewline
 &  &  &  &  &  &  &  & $\arrowvert H-2\rightarrow L+1\rangle(0.4246)$\tabularnewline
 &  &  &  &  &  &  &  & \tabularnewline
V\textsubscript{$xy$} &  & \textsuperscript{1}A\textsubscript{1} &  & 6.54 &  & 11.628 &  & $\arrowvert H-2\rightarrow L+3\rangle(0.3241)$\tabularnewline
 &  &  &  &  &  &  &  & $\arrowvert H-3\rightarrow L+2\rangle(0.3241)$\tabularnewline
 &  & \textsuperscript{1}B\textsubscript{2} &  & 6.64 &  & 7.875 &  & $\arrowvert H-4\rightarrow L+2\rangle(0.3611)$\tabularnewline
 &  &  &  &  &  &  &  & $\arrowvert H-2\rightarrow L+4\rangle(0.3611)$\tabularnewline
 &  &  &  &  &  &  &  & \tabularnewline
VI\textsubscript{$xy$} &  & \textsuperscript{1}A\textsubscript{1} &  & 7.03 &  & 2.237 &  & $\arrowvert H-3\rightarrow L+4\rangle(0.2981)$\tabularnewline
 &  &  &  &  &  &  &  & $\arrowvert H-4\rightarrow L+3\rangle(0.2981)$\tabularnewline
 &  & \textsuperscript{1}B\textsubscript{2} &  & 7.17 &  & 5.022 &  & $\arrowvert H-5\rightarrow L+5\rangle(0.3502)$\tabularnewline
 &  &  &  &  &  &  &  & $\arrowvert H-3\rightarrow L+3\rangle(0.2922)$\tabularnewline
 &  & \textsuperscript{1}B\textsubscript{2} &  & 7.39 &  & 1.165 &  & $\arrowvert H-4\rightarrow L+6\rangle(0.2894)$\tabularnewline
 &  &  &  &  &  &  &  & $\arrowvert H-6\rightarrow L+4\rangle(0.2894)$\tabularnewline
\hline 
\end{tabular}
\end{table}

\begin{table}[H]
\caption{Many-particle wave functions of the excited states contributing to
the peaks in the linear optical absorption spectra of naphtho{[}2,3a{]}coronene
(C\protect\textsubscript{2v}) {[}Fig. 1c, main article{]} computed
using the screened parameters in the PPP semi-empirical method and
MRSDCI methodology. In the 'Peak' column, $x$ ($y$) indicates that
the polarization of the absorbed photon is along $x$ ($y$) direction,
while $xy$ denotes mixed polarization. The excited state symmetries
of the corresponding peaks are also presented below. $f$ (= $\frac{2\times m_{e}}{3\times\hbar^{2}}(E)\sum_{j}|\langle e|O_{j}|R\rangle|^{2}$)
indicate the oscillator strength for a particular transition where
$E$, $|e\rangle$, $|R\rangle$ and $O_{j}$ representing the peak
energy, excited states of the corresponding peak, reference state
and electric dipole operator for different Cartesian components, respectively.
In the last column, the fractional numbers inside the first bracket
represent the contributing coefficients to the CI wave functions for
each corresponding configuration.}

\begin{tabular}{ccccccccc}
\hline 
Peak & \hspace{1.7cm} & Symmetry & \hspace{1.7cm} & E (eV) & \hspace{1.7cm} & $f$ & \hspace{1.7cm} & Wave function\tabularnewline
\hline 
\hline 
 &  &  &  &  &  &  &  & \tabularnewline
I\textsubscript{$y$} &  & \textsuperscript{1}B\textsubscript{2} &  & 2.93 &  & 2.651 &  & $\arrowvert H\rightarrow L\rangle(0.8638)$\tabularnewline
 &  &  &  &  &  &  &  & $\arrowvert H-1\rightarrow L+1\rangle(0.0667)$\tabularnewline
 &  &  &  &  &  &  &  & \tabularnewline
II\textsubscript{$x$} &  & \textsuperscript{1}A\textsubscript{1} &  & 3.48 &  & 17.442 &  & $\arrowvert H\rightarrow L+1\rangle(0.6118)$\tabularnewline
 &  &  &  &  &  &  &  & $\arrowvert H-1\rightarrow L\rangle(0.6118)$\tabularnewline
 &  &  &  &  &  &  &  & \tabularnewline
III\textsubscript{$y$} &  & \textsuperscript{1}B\textsubscript{2} &  & 3.88 &  & 5.994 &  & $\arrowvert H-1\rightarrow L+1\rangle(0.7912)$\tabularnewline
 &  &  &  &  &  &  &  & $\arrowvert H-2\rightarrow L\rangle(0.2232)$\tabularnewline
 &  &  &  &  &  &  &  & \tabularnewline
IV\textsubscript{$xy$} &  & \textsuperscript{1}A\textsubscript{1} &  & 4.33 &  & 1.119 &  & $\arrowvert H-1\rightarrow L+2\rangle(0.5879)$\tabularnewline
 &  &  &  &  &  &  &  & $\arrowvert H-2\rightarrow L+1\rangle(0.5879)$\tabularnewline
 &  & \textsuperscript{1}B\textsubscript{2} &  & 4.45 &  & 3.117 &  & $\arrowvert H-2\rightarrow L+2\rangle(0.8243)$\tabularnewline
 &  &  &  &  &  &  &  & $\arrowvert H\rightarrow L+6\rangle(0.1066)$\tabularnewline
 &  & \textsuperscript{1}A\textsubscript{1} &  & 4.48 &  & 1.543 &  & $\arrowvert H\rightarrow L+3\rangle(0.5364)$\tabularnewline
 &  &  &  &  &  &  &  & $\arrowvert H-3\rightarrow L\rangle(0.5364)$\tabularnewline
 &  &  &  &  &  &  &  & \tabularnewline
V\textsubscript{$xy$} &  & \textsuperscript{1}B\textsubscript{2} &  & 5.16 &  & 1.624 &  & $\arrowvert H-2\rightarrow L+4\rangle(0.5825)$\tabularnewline
 &  &  &  &  &  &  &  & $\arrowvert H-4\rightarrow L+2\rangle(0.5825)$\tabularnewline
 &  & \textsuperscript{1}A\textsubscript{1} &  & 5.21 &  & 3.558 &  & $\arrowvert H-2\rightarrow L+3\rangle(0.5182)$\tabularnewline
 &  &  &  &  &  &  &  & $\arrowvert H-3\rightarrow L+2\rangle(0.5182)$\tabularnewline
 &  & \textsuperscript{1}A\textsubscript{1} &  & 5.38 &  & 0.918 &  & $\arrowvert H\rightarrow L+7\rangle(0.4545)$\tabularnewline
 &  &  &  &  &  &  &  & $\arrowvert H-7\rightarrow L\rangle(0.4545)$\tabularnewline
 &  &  &  &  &  &  &  & \tabularnewline
VI\textsubscript{$x$} &  & \textsuperscript{1}A\textsubscript{1} &  & 5.62 &  & 2.090 &  & $\arrowvert H-1\rightarrow L+6\rangle(0.4026)$\tabularnewline
 &  &  &  &  &  &  &  & $\arrowvert H-6\rightarrow L+1\rangle(0.4026)$\tabularnewline
 &  &  &  &  &  &  &  & \tabularnewline
VII\textsubscript{$xy$} &  & \textsuperscript{1}A\textsubscript{1} &  & 5.88 &  & 1.722 &  & $\arrowvert H\rightarrow L+9\rangle(0.4545)$\tabularnewline
 &  &  &  &  &  &  &  & $\arrowvert H-9\rightarrow L\rangle(0.4545)$\tabularnewline
 &  & \textsuperscript{1}B\textsubscript{2} &  & 6.07 &  & 3.035 &  & $\arrowvert H-3\rightarrow L+3\rangle(0.4929)$\tabularnewline
 &  &  &  &  &  &  &  & $\arrowvert H-2\rightarrow L;H-1\rightarrow L\rangle(0.2906)$\tabularnewline
 &  &  &  &  &  &  &  & \tabularnewline
VIII\textsubscript{$xy$} &  & \textsuperscript{1}A\textsubscript{1} &  & 6.28 &  & 2.475 &  & $\arrowvert H-4\rightarrow L+5\rangle(0.2986)$\tabularnewline
 &  &  &  &  &  &  &  & $\arrowvert H-5\rightarrow L+4\rangle(0.2986)$\tabularnewline
 &  & \textsuperscript{1}B\textsubscript{2} &  & 6.33 &  & 0.730 &  & $\arrowvert H-2\rightarrow L+8\rangle(0.4299)$\tabularnewline
 &  &  &  &  &  &  &  & $\arrowvert H-8\rightarrow L+2\rangle(0.4299)$\tabularnewline
 &  &  &  &  &  &  &  & \tabularnewline
IX\textsubscript{$xy$} &  & \textsuperscript{1}B\textsubscript{2} &  & 6.84 &  & 0.302 &  & $\arrowvert H-6\rightarrow L+6\rangle(0.7206)$\tabularnewline
 &  &  &  &  &  &  &  & $\arrowvert H-11\rightarrow L+1\rangle(0.1724)$\tabularnewline
 &  & \textsuperscript{1}A\textsubscript{1} &  & 7.03 &  & 0.662 &  & $\arrowvert H-8\rightarrow L+3\rangle(0.5536)$\tabularnewline
 &  &  &  &  &  &  &  & $\arrowvert H-3\rightarrow L+8\rangle(0.5536)$\tabularnewline
\hline 
\end{tabular}
\end{table}

\begin{table}[H]
\caption{Many-particle wave functions of the excited states contributing to
the peaks in the linear optical absorption spectra of naphtho{[}2,3a{]}coronene
(C\protect\textsubscript{2v}) {[}Fig. 1c, main article{]} computed
using the standard parameters in the PPP semi-empirical method and
MRSDCI methodology. In the 'Peak' column, $x$ ($y$) indicates that
the polarization of the absorbed photon is along $x$ ($y$) direction,
while $xy$ denotes mixed polarization. The excited state symmetries
of the corresponding peaks are also presented below. $f$ (= $\frac{2\times m_{e}}{3\times\hbar^{2}}(E)\sum_{j}|\langle e|O_{j}|R\rangle|^{2}$)
indicate the oscillator strength for a particular transition where
$E$, $|e\rangle$, $|R\rangle$ and $O_{j}$ representing the peak
energy, excited states of the corresponding peak, reference state
and electric dipole operator for different Cartesian components, respectively.
In the last column, the fractional numbers inside the first bracket
represent the contributing coefficients to the CI wave functions for
each corresponding configuration.}

\begin{tabular}{ccccccccc}
\hline 
Peak & \hspace{1.7cm} & Symmetry & \hspace{1.7cm} & E (eV) & \hspace{1.7cm} & $f$ & \hspace{1.7cm} & Wave function\tabularnewline
\hline 
\hline 
 &  &  &  &  &  &  &  & \tabularnewline
I\textsubscript{$y$} &  & \textsuperscript{1}B\textsubscript{2} &  & 3.32 &  & 1.187 &  & $\arrowvert H\rightarrow L\rangle(0.7861)$\tabularnewline
 &  &  &  &  &  &  &  & $\arrowvert H-1\rightarrow L+1\rangle(0.3266)$\tabularnewline
 &  &  &  &  &  &  &  & \tabularnewline
II\textsubscript{$x$} &  & \textsuperscript{1}A\textsubscript{1} &  & 4.19 &  & 15.692 &  & $\arrowvert H-1\rightarrow L\rangle(0.5783)$\tabularnewline
 &  &  &  &  &  &  &  & $\arrowvert H\rightarrow L+1\rangle(0.5783)$\tabularnewline
 &  &  &  &  &  &  &  & \tabularnewline
III\textsubscript{$y$} &  & \textsuperscript{1}B\textsubscript{2} &  & 4.50 &  & 3.826 &  & $\arrowvert H-1\rightarrow L+1\rangle(0.6668)$\tabularnewline
 &  &  &  &  &  &  &  & $\arrowvert H-2\rightarrow L+2\rangle(0.2850)$\tabularnewline
 &  &  &  &  &  &  &  & \tabularnewline
IV\textsubscript{$xy$} &  & \textsuperscript{1}B\textsubscript{2} &  & 5.22 &  & 3.403 &  & $\arrowvert H-2\rightarrow L+2\rangle(0.5454)$\tabularnewline
 &  &  &  &  &  &  &  & $\arrowvert H-1\rightarrow L+3\rangle(0.2414)$\tabularnewline
 &  & \textsuperscript{1}A\textsubscript{1} &  & 5.23 &  & 1.655 &  & $\arrowvert H-2\rightarrow L+1\rangle(0.4793)$\tabularnewline
 &  &  &  &  &  &  &  & $\arrowvert H-1\rightarrow L+2\rangle(0.4793)$\tabularnewline
 &  & \textsuperscript{1}A\textsubscript{1} &  & 5.39 &  & 2.008 &  & $\arrowvert H\rightarrow L+3\rangle(0.4040)$\tabularnewline
 &  &  &  &  &  &  &  & $\arrowvert H-3\rightarrow L\rangle(0.4040)$\tabularnewline
 &  &  &  &  &  &  &  & \tabularnewline
V\textsubscript{$x$} &  & \textsuperscript{1}A\textsubscript{1} &  & 5.96 &  & 1.833 &  & $\arrowvert H-1\rightarrow L+4\rangle(0.4476)$\tabularnewline
 &  &  &  &  &  &  &  & $\arrowvert H-4\rightarrow L+1\rangle(0.4476)$\tabularnewline
 &  &  &  &  &  &  &  & \tabularnewline
VI\textsubscript{$xy$} &  & \textsuperscript{1}A\textsubscript{1} &  & 6.24 &  & 3.457 &  & $\arrowvert H-3\rightarrow L+2\rangle(0.4533)$\tabularnewline
 &  &  &  &  &  &  &  & $\arrowvert H-2\rightarrow L+3\rangle(0.4533)$\tabularnewline
 &  & \textsuperscript{1}B\textsubscript{2} &  & 6.34 &  & 2.579 &  & $\arrowvert H-4\rightarrow L+2\rangle(0.3554)$\tabularnewline
 &  &  &  &  &  &  &  & $\arrowvert H-2\rightarrow L+4\rangle(0.3554)$\tabularnewline
 &  &  &  &  &  &  &  & \tabularnewline
VII\textsubscript{$x$} &  & \textsuperscript{1}A\textsubscript{1} &  & 6.59 &  & 2.025 &  & $\arrowvert H-2\rightarrow L+5\rangle(0.4343)$\tabularnewline
 &  &  &  &  &  &  &  & $\arrowvert H-5\rightarrow L+2\rangle(0.4343)$\tabularnewline
 &  &  &  &  &  &  &  & \tabularnewline
VIII\textsubscript{$xy$} &  & \textsuperscript{1}A\textsubscript{1} &  & 6.85 &  & 0.772 &  & $\arrowvert H-9\rightarrow L\rangle(0.2840)$\tabularnewline
 &  &  &  &  &  &  &  & $\arrowvert H\rightarrow L+9\rangle(0.2840)$\tabularnewline
 &  & \textsuperscript{1}B\textsubscript{2} &  & 6.88 &  & 2.852 &  & $\arrowvert H-4\rightarrow L+4\rangle(0.4994)$\tabularnewline
 &  &  &  &  &  &  &  & $\arrowvert H-5\rightarrow L+5\rangle(0.2536)$\tabularnewline
 &  &  &  &  &  &  &  & \tabularnewline
IX\textsubscript{$xy$} &  & \textsuperscript{1}B\textsubscript{2} &  & 7.04 &  & 1.970 &  & $\arrowvert H-3\rightarrow L+3\rangle(0.4102)$\tabularnewline
 &  &  &  &  &  &  &  & $\arrowvert H-7\rightarrow L+1\rangle(0.2334)$\tabularnewline
 &  & \textsuperscript{1}A\textsubscript{1} &  & 7.09 &  & 1.101 &  & $\arrowvert H-3\rightarrow L+4\rangle(0.3228)$\tabularnewline
 &  &  &  &  &  &  &  & $\arrowvert H-4\rightarrow L+3\rangle(0.3228)$\tabularnewline
 &  &  &  &  &  &  &  & \tabularnewline
X\textsubscript{$xy$} &  & \textsuperscript{1}A\textsubscript{1} &  & 7.34 &  & 3.468 &  & $\arrowvert H-3\rightarrow L+6\rangle(0.3729)$\tabularnewline
 &  &  &  &  &  &  &  & $\arrowvert H-6\rightarrow L+3\rangle(0.3729)$\tabularnewline
 &  & \textsuperscript{1}B\textsubscript{2} &  & 7.40 &  & 3.074 &  & $\arrowvert H-9\rightarrow L+1\rangle(0.3128)$\tabularnewline
 &  &  &  &  &  &  &  & $\arrowvert H-1\rightarrow L+9\rangle(0.3128)$\tabularnewline
\hline 
\end{tabular}
\end{table}

\begin{table}[H]
\caption{Many-particle wave functions of the excited states contributing to
the peaks in the linear optical absorption spectra of anthra{[}2,3a{]}coronene
(C\protect\textsubscript{2v}) {[}Fig. 1d, main article{]} computed
using the screened parameters in the PPP semi-empirical method and
MRSDCI methodology. In the 'Peak' column, $x$ ($y$) indicates that
the polarization of the absorbed photon is along $x$ ($y$) direction,
while $xy$ denotes mixed polarization. The excited state symmetries
of the corresponding peaks are also presented below. $f$ (= $\frac{2\times m_{e}}{3\times\hbar^{2}}(E)\sum_{j}|\langle e|O_{j}|R\rangle|^{2}$)
indicate the oscillator strength for a particular transition where
$E$, $|e\rangle$, $|R\rangle$ and $O_{j}$ representing the peak
energy, excited states of the corresponding peak, reference state
and electric dipole operator for different Cartesian components, respectively.
In the last column, the fractional numbers inside the first bracket
represent the contributing coefficients to the CI wave functions for
each corresponding configuration.}

\begin{tabular}{ccccccccc}
\hline 
Peak & \hspace{1.7cm} & Symmetry & \hspace{1.7cm} & E (eV) & \hspace{1.7cm} & $f$ & \hspace{1.7cm} & Wave function\tabularnewline
\hline 
\hline 
 &  &  &  &  &  &  &  & \tabularnewline
I\textsubscript{$y$} &  & \textsuperscript{1}B\textsubscript{2} &  & 2.70 &  & 1.964 &  & $\arrowvert H\rightarrow L\rangle(0.8495)$\tabularnewline
 &  &  &  &  &  &  &  & $\arrowvert H\rightarrow L+1\rangle(0.0707)$\tabularnewline
 &  &  &  &  &  &  &  & \tabularnewline
II\textsubscript{$xy$} &  & \textsuperscript{1}B\textsubscript{2} &  & 3.32 &  & 0.232 &  & $\arrowvert H\rightarrow L+1\rangle(0.5981)$\tabularnewline
 &  &  &  &  &  &  &  & $\arrowvert H-1\rightarrow L\rangle(0.5981)$\tabularnewline
 &  & \textsuperscript{1}A\textsubscript{1} &  & 3.41 &  & 17.298 &  & $\arrowvert H\rightarrow L+2\rangle(0.6023)$\tabularnewline
 &  &  &  &  &  &  &  & $\arrowvert H-2\rightarrow L\rangle(0.6023)$\tabularnewline
 &  &  &  &  &  &  &  & \tabularnewline
III\textsubscript{$xy$} &  & \textsuperscript{1}B\textsubscript{2} &  & 4.04 &  & 8.611 &  & $\arrowvert H-2\rightarrow L+2\rangle(0.6536)$\tabularnewline
 &  &  &  &  &  &  &  & $\arrowvert H-1\rightarrow L+1\rangle(0.4869)$\tabularnewline
 &  & \textsuperscript{1}A\textsubscript{1} &  & 4.16 &  & 4.877 &  & $\arrowvert H-2\rightarrow L+1\rangle(0.5713)$\tabularnewline
 &  &  &  &  &  &  &  & $\arrowvert H-1\rightarrow L+2\rangle(0.5713)$\tabularnewline
 &  &  &  &  &  &  &  & \tabularnewline
IV\textsubscript{$xy$} &  & \textsuperscript{1}B\textsubscript{2} &  & 4.30 &  & 1.698 &  & $\arrowvert H-5\rightarrow L\rangle(0.5545)$\tabularnewline
 &  &  &  &  &  &  &  & $\arrowvert H\rightarrow L+5\rangle(0.5545)$\tabularnewline
 &  & \textsuperscript{1}A\textsubscript{1} &  & 4.40 &  & 3.663 &  & $\arrowvert H-4\rightarrow L\rangle(0.5701)$\tabularnewline
 &  &  &  &  &  &  &  & $\arrowvert H\rightarrow L+4\rangle(0.5701)$\tabularnewline
 &  &  &  &  &  &  &  & \tabularnewline
V\textsubscript{$xy$} &  & \textsuperscript{1}A\textsubscript{1} &  & 4.98 &  & 0.877 &  & $\arrowvert H-4\rightarrow L+1\rangle(0.3220)$\tabularnewline
 &  &  &  &  &  &  &  & $\arrowvert H-1\rightarrow L+4\rangle(0.3220)$\tabularnewline
 &  & \textsuperscript{1}B\textsubscript{2} &  & 5.10 &  & 0.342 &  & $\arrowvert H-7\rightarrow L\rangle(0.4830)$\tabularnewline
 &  &  &  &  &  &  &  & $\arrowvert H\rightarrow L+7\rangle(0.4830)$\tabularnewline
 &  & \textsuperscript{1}A\textsubscript{1} &  & 5.12 &  & 0.851 &  & $\arrowvert H\rightarrow L+8\rangle(0.3172)$\tabularnewline
 &  &  &  &  &  &  &  & $\arrowvert H-8\rightarrow L\rangle(0.3172)$\tabularnewline
 &  &  &  &  &  &  &  & \tabularnewline
VI\textsubscript{$x$} &  & \textsuperscript{1}A\textsubscript{1} &  & 5.33 &  & 1.948 &  & $\arrowvert H-8\rightarrow L\rangle(0.4022)$\tabularnewline
 &  &  &  &  &  &  &  & $\arrowvert H\rightarrow L+8\rangle(0.4022)$\tabularnewline
 &  &  &  &  &  &  &  & \tabularnewline
VII\textsubscript{$xy$} &  & \textsuperscript{1}A\textsubscript{1} &  & 5.90 &  & 2.248 &  & $\arrowvert H-10\rightarrow L\rangle(0.4076)$\tabularnewline
 &  &  &  &  &  &  &  & $\arrowvert H\rightarrow L+10\rangle(0.4076)$\tabularnewline
 &  & \textsuperscript{1}B\textsubscript{2} &  & 6.08 &  & 2.730 &  & $\arrowvert H-3\rightarrow L+3\rangle(0.4667)$\tabularnewline
 &  &  &  &  &  &  &  & $\arrowvert H-5\rightarrow L+5\rangle(0.2173)$\tabularnewline
 &  &  &  &  &  &  &  & \tabularnewline
VIII\textsubscript{$xy$} &  & \textsuperscript{1}A\textsubscript{1} &  & 6.50 &  & 1.508 &  & $\arrowvert H\rightarrow L+1;H\rightarrow L+3\rangle(0.3760)$\tabularnewline
 &  &  &  &  &  &  &  & $\arrowvert H-1\rightarrow L;H-3\rightarrow L\rangle(0.3760)$\tabularnewline
 &  & \textsuperscript{1}B\textsubscript{2} &  & 6.50 &  & 1.338 &  & $\arrowvert H-6\rightarrow L+6\rangle(0.4974)$\tabularnewline
 &  &  &  &  &  &  &  & $\arrowvert H-5\rightarrow L+5\rangle(0.4342)$\tabularnewline
 &  &  &  &  &  &  &  & \tabularnewline
IX\textsubscript{$xy$} &  & \textsuperscript{1}B\textsubscript{2} &  & 6.78 &  & 3.160 &  & $\arrowvert H-7\rightarrow L+5\rangle(0.5244)$\tabularnewline
 &  &  &  &  &  &  &  & $\arrowvert H-5\rightarrow L+7\rangle(0.5244)$\tabularnewline
\hline 
\end{tabular}
\end{table}

\begin{table}[H]
\caption{Many-particle wave functions of the excited states contributing to
the peaks in the linear optical absorption spectra of anthra{[}2,3a{]}coronene
(C\protect\textsubscript{2v}) {[}Fig. 1d, main article{]} computed
using the standard parameters in the PPP semi-empirical method and
MRSDCI methodology. In the 'Peak' column, $x$ ($y$) indicates that
the polarization of the absorbed photon is along $x$ ($y$) direction,
while $xy$ denotes mixed polarization. The excited state symmetries
of the corresponding peaks are also presented below. $f$ (= $\frac{2\times m_{e}}{3\times\hbar^{2}}(E)\sum_{j}|\langle e|O_{j}|R\rangle|^{2}$)
indicate the oscillator strength for a particular transition where
$E$, $|e\rangle$, $|R\rangle$ and $O_{j}$ representing the peak
energy, excited states of the corresponding peak, reference state
and electric dipole operator for different Cartesian components, respectively.
In the last column, the fractional numbers inside the first bracket
represent the contributing coefficients to the CI wave functions for
each corresponding configuration.}

\begin{tabular}{ccccccccc}
\hline 
Peak & \hspace{1.7cm} & Symmetry & \hspace{1.7cm} & E (eV) & \hspace{1.7cm} & $f$ & \hspace{1.7cm} & Wave function\tabularnewline
\hline 
\hline 
 &  &  &  &  &  &  &  & \tabularnewline
I\textsubscript{$y$} &  & \textsuperscript{1}B\textsubscript{2} &  & 3.10 &  & 1.561 &  & $\arrowvert H\rightarrow L\rangle(0.8104)$\tabularnewline
 &  &  &  &  &  &  &  & $\arrowvert H-2\rightarrow L+2\rangle(0.1950)$\tabularnewline
 &  &  &  &  &  &  &  & \tabularnewline
II\textsubscript{$x$} &  & \textsuperscript{1}A\textsubscript{1} &  & 4.10 &  & 19.168 &  & $\arrowvert H-2\rightarrow L\rangle(0.5721)$\tabularnewline
 &  &  &  &  &  &  &  & $\arrowvert H\rightarrow L+2\rangle(0.5721)$\tabularnewline
 &  &  &  &  &  &  &  & \tabularnewline
III\textsubscript{$y$} &  & \textsuperscript{1}B\textsubscript{2} &  & 4.71 &  & 2.906 &  & $\arrowvert H\rightarrow L+4\rangle(0.3390)$\tabularnewline
 &  &  &  &  &  &  &  & $\arrowvert H-4\rightarrow L\rangle(0.3390)$\tabularnewline
 &  &  &  &  &  &  &  & \tabularnewline
IV\textsubscript{$xy$} &  & \textsuperscript{1}B\textsubscript{2} &  & 4.99 &  & 2.144 &  & $\arrowvert H-1\rightarrow L+1\rangle(0.4079)$\tabularnewline
 &  &  &  &  &  &  &  & $\arrowvert H-2\rightarrow L+3\rangle(0.3218)$\tabularnewline
 &  & \textsuperscript{1}A\textsubscript{1} &  & 5.12 &  & 4.876 &  & $\arrowvert H-2\rightarrow L+1\rangle(0.4783)$\tabularnewline
 &  &  &  &  &  &  &  & $\arrowvert H-1\rightarrow L+2\rangle(0.4783)$\tabularnewline
 &  & \textsuperscript{1}B\textsubscript{2} &  & 5.15 &  & 1.693 &  & $\arrowvert H-1\rightarrow L+1\rangle(0.3198)$\tabularnewline
 &  &  &  &  &  &  &  & $\arrowvert H-1\rightarrow L+4\rangle(0.2917)$\tabularnewline
 &  &  &  &  &  &  &  & \tabularnewline
V\textsubscript{$xy$} &  & \textsuperscript{1}A\textsubscript{1} &  & 5.90 &  & 2.332 &  & $\arrowvert H-6\rightarrow L\rangle(0.3121)$\tabularnewline
 &  &  &  &  &  &  &  & $\arrowvert H\rightarrow L+6\rangle(0.3121)$\tabularnewline
 &  & \textsuperscript{1}B\textsubscript{2} &  & 5.94 &  & 0.740 &  & $\arrowvert H-5\rightarrow L+1\rangle(0.3600)$\tabularnewline
 &  &  &  &  &  &  &  & $\arrowvert H-1\rightarrow L+5\rangle(0.3600)$\tabularnewline
 &  &  &  &  &  &  &  & \tabularnewline
VI\textsubscript{$x$} &  & \textsuperscript{1}A\textsubscript{1} &  & 6.24 &  & 1.454 &  & $\arrowvert H-8\rightarrow L\rangle(0.3152)$\tabularnewline
 &  &  &  &  &  &  &  & $\arrowvert H\rightarrow L+8\rangle(0.3152)$\tabularnewline
 &  &  &  &  &  &  &  & \tabularnewline
VII\textsubscript{$xy$} &  & \textsuperscript{1}A\textsubscript{1} &  & 6.90 &  & 1.418 &  & $\arrowvert H-7\rightarrow L+2\rangle(0.2677)$\tabularnewline
 &  &  &  &  &  &  &  & $\arrowvert H-2\rightarrow L+7\rangle(0.2677)$\tabularnewline
 &  & \textsuperscript{1}B\textsubscript{2} &  & 6.93 &  & 2.132 &  & $\arrowvert H-5\rightarrow L+5\rangle(0.2927)$\tabularnewline
 &  &  &  &  &  &  &  & $\arrowvert H-6\rightarrow L+6\rangle(0.2493)$\tabularnewline
 &  &  &  &  &  &  &  & \tabularnewline
VIII\textsubscript{$xy$} &  & \textsuperscript{1}B\textsubscript{2} &  & 7.06 &  & 1.157 &  & $\arrowvert H-6\rightarrow L+6\rangle(0.2126)$\tabularnewline
 &  &  &  &  &  &  &  & $\arrowvert H-4\rightarrow L+4\rangle(0.2085)$\tabularnewline
 &  & \textsuperscript{1}A\textsubscript{1} &  & 7.10 &  & 2.757 &  & $\arrowvert H-8\rightarrow L+1\rangle(0.3200)$\tabularnewline
 &  &  &  &  &  &  &  & $\arrowvert H-1\rightarrow L+8\rangle(0.3200)$\tabularnewline
 &  &  &  &  &  &  &  & \tabularnewline
IX\textsubscript{$xy$} &  & \textsuperscript{1}A\textsubscript{1} &  & 7.29 &  & 1.751 &  & $\arrowvert H-5\rightarrow L+3\rangle(0.4112)$\tabularnewline
 &  &  &  &  &  &  &  & $\arrowvert H-3\rightarrow L+5\rangle(0.4112)$\tabularnewline
 &  & \textsuperscript{1}B\textsubscript{2} &  & 7.30 &  & 4.616 &  & $\arrowvert H-3\rightarrow L+3\rangle(0.3333)$\tabularnewline
 &  &  &  &  &  &  &  & $\arrowvert H-4\rightarrow L+4\rangle(0.3218)$\tabularnewline
\hline 
\end{tabular}
\end{table}

\begin{table}[H]
\caption{Many-particle wave functions of the excited states contributing to
the peaks in the linear optical absorption spectra of naphtho{[}8,1,2-abc{]}coronene
(C\protect\textsubscript{s}) {[}Fig. 1e, main article{]} computed
using the screened parameters in the PPP semi-empirical method and
MRSDCI methodology. In the 'Peak' column, $xy$ denotes mixed polarization.
The excited state symmetries of the corresponding peaks are also presented
below. $f$ (= $\frac{2\times m_{e}}{3\times\hbar^{2}}(E)\sum_{j}|\langle e|O_{j}|R\rangle|^{2}$)
indicate the oscillator strength for a particular transition where
$E$, $|e\rangle$, $|R\rangle$ and $O_{j}$ representing the peak
energy, excited states of the corresponding peak, reference state
and electric dipole operator for different Cartesian components, respectively.
In the last column, the fractional numbers inside the first bracket
represent the contributing coefficients to the CI wave functions for
each corresponding configuration.}

\begin{tabular}{ccccccccc}
\hline 
Peak & \hspace{1.7cm} & Symmetry & \hspace{1.7cm} & E (eV) & \hspace{1.7cm} & $f$ & \hspace{1.7cm} & Wave function\tabularnewline
\hline 
\hline 
 &  &  &  &  &  &  &  & \tabularnewline
I\textsubscript{$xy$} &  & \textsuperscript{1}A &  & 2.95 &  & 4.123 &  & $\arrowvert H\rightarrow L\rangle(0.8665)$\tabularnewline
 &  &  &  &  &  &  &  & $\arrowvert H-1\rightarrow L+1\rangle(0.0766)$\tabularnewline
 &  &  &  &  &  &  &  & \tabularnewline
II\textsubscript{$xy$} &  & \textsuperscript{1}A &  & 3.43 &  & 12.024 &  & $\arrowvert H-1\rightarrow L\rangle(0.6096)$\tabularnewline
 &  &  &  &  &  &  &  & $\arrowvert H\rightarrow L+1\rangle(0.6096)$\tabularnewline
 &  &  &  &  &  &  &  & \tabularnewline
III\textsubscript{$xy$} &  & \textsuperscript{1}A &  & 3.89 &  & 3.940 &  & $\arrowvert H-1\rightarrow L+1\rangle(0.6975)$\tabularnewline
 &  &  &  &  &  &  &  & $\arrowvert H-2\rightarrow L\rangle(0.3205)$\tabularnewline
 &  &  &  &  &  &  &  & \tabularnewline
IV\textsubscript{$xy$} &  & \textsuperscript{1}A &  & 4.56 &  & 1.023 &  & $\arrowvert H-2\rightarrow L+1\rangle(0.4278)$\tabularnewline
 &  &  &  &  &  &  &  & $\arrowvert H-1\rightarrow L+2\rangle(0.4278)$\tabularnewline
 &  &  &  & 4.72 &  & 0.841 &  & $\arrowvert H-3\rightarrow L+1\rangle(0.5139)$\tabularnewline
 &  &  &  &  &  &  &  & $\arrowvert H-1\rightarrow L+3\rangle(0.5139)$\tabularnewline
 &  &  &  &  &  &  &  & \tabularnewline
V\textsubscript{$xy$} &  & \textsuperscript{1}A &  & 5.03 &  & 1.974 &  & $\arrowvert H-4\rightarrow L+1\rangle(0.6446)$\tabularnewline
 &  &  &  &  &  &  &  & $\arrowvert H-1\rightarrow L+4\rangle(0.2761)$\tabularnewline
 &  &  &  &  &  &  &  & \tabularnewline
VI\textsubscript{$xy$} &  & \textsuperscript{1}A &  & 5.54 &  & 1.877 &  & $\arrowvert H-6\rightarrow L+1\rangle(0.4299)$\tabularnewline
 &  &  &  &  &  &  &  & $\arrowvert H-1\rightarrow L+6\rangle(0.4299)$\tabularnewline
 &  &  &  &  &  &  &  & \tabularnewline
VII\textsubscript{$xy$} &  & \textsuperscript{1}A &  & 5.72 &  & 2.535 &  & $\arrowvert H-8\rightarrow L\rangle(0.4521)$\tabularnewline
 &  &  &  &  &  &  &  & $\arrowvert H\rightarrow L+8\rangle(0.4521)$\tabularnewline
 &  &  &  &  &  &  &  & \tabularnewline
VIII\textsubscript{$xy$} &  & \textsuperscript{1}A &  & 5.91 &  & 4.752 &  & $\arrowvert H-3\rightarrow L+4\rangle(0.3762)$\tabularnewline
 &  &  &  &  &  &  &  & $\arrowvert H-4\rightarrow L+3\rangle(0.3762)$\tabularnewline
 &  &  &  &  &  &  &  & \tabularnewline
IX\textsubscript{$xy$} &  & \textsuperscript{1}A &  & 6.39 &  & 2.840 &  & $\arrowvert H-1\rightarrow L+8\rangle(0.2899)$\tabularnewline
 &  &  &  &  &  &  &  & $\arrowvert H-8\rightarrow L+1\rangle(0.2899)$\tabularnewline
\hline 
\end{tabular}
\end{table}

\begin{table}[H]
\caption{Many-particle wave functions of the excited states contributing to
the peaks in the linear optical absorption spectra of naphtho{[}8,1,2-abc{]}coronene
(C\protect\textsubscript{s}) {[}Fig. 1e, main article{]} computed
using the standard parameters in the PPP semi-empirical method and
MRSDCI methodology. In the 'Peak' column, $xy$ denotes mixed polarization.
The excited state symmetries of the corresponding peaks are also presented
below. $f$ (= $\frac{2\times m_{e}}{3\times\hbar^{2}}(E)\sum_{j}|\langle e|O_{j}|R\rangle|^{2}$)
indicate the oscillator strength for a particular transition where
$E$, $|e\rangle$, $|R\rangle$ and $O_{j}$ representing the peak
energy, excited states of the corresponding peak, reference state
and electric dipole operator for different Cartesian components, respectively.
In the last column, the fractional numbers inside the first bracket
represent the contributing coefficients to the CI wave functions for
each corresponding configuration.}

\begin{tabular}{ccccccccc}
\hline 
Peak & \hspace{1.7cm} & Symmetry & \hspace{1.7cm} & E (eV) & \hspace{1.7cm} & $f$ & \hspace{1.7cm} & Wave function\tabularnewline
\hline 
\hline 
 &  &  &  &  &  &  &  & \tabularnewline
I\textsubscript{$xy$} &  & \textsuperscript{1}A &  & 3.23 &  & 1.810 &  & $\arrowvert H\rightarrow L\rangle(0.7986)$\tabularnewline
 &  &  &  &  &  &  &  & $\arrowvert H-1\rightarrow L+1\rangle(0.3175)$\tabularnewline
 &  &  &  &  &  &  &  & \tabularnewline
II\textsubscript{$xy$} &  & \textsuperscript{1}A &  & 3.98 &  & 4.083 &  & $\arrowvert H-2\rightarrow L\rangle(0.4359)$\tabularnewline
 &  &  &  &  &  &  &  & $\arrowvert H\rightarrow L+2\rangle(0.4359)$\tabularnewline
 &  &  &  &  &  &  &  & \tabularnewline
III\textsubscript{$xy$} &  & \textsuperscript{1}A &  & 4.21 &  & 6.136 &  & $\arrowvert H-1\rightarrow L\rangle(0.4236)$\tabularnewline
 &  &  &  &  &  &  &  & $\arrowvert H\rightarrow L+1\rangle(0.4236)$\tabularnewline
 &  &  &  &  &  &  &  & \tabularnewline
IV\textsubscript{$xy$} &  & \textsuperscript{1}A &  & 4.37 &  & 4.586 &  & $\arrowvert H-1\rightarrow L+1\rangle(0.6742)$\tabularnewline
 &  &  &  &  &  &  &  & $\arrowvert H\rightarrow L\rangle(0.2422)$\tabularnewline
 &  &  &  &  &  &  &  & \tabularnewline
V\textsubscript{$xy$} &  & \textsuperscript{1}A &  & 5.04 &  & 0.741 &  & $\arrowvert H-6\rightarrow L\rangle(0.3924)$\tabularnewline
 &  &  &  &  &  &  &  & $\arrowvert H\rightarrow L+6\rangle(0.3924)$\tabularnewline
 &  &  &  &  &  &  &  & \tabularnewline
VI\textsubscript{$xy$} &  & \textsuperscript{1}A &  & 5.41 &  & 1.648 &  & $\arrowvert H-1\rightarrow L+2\rangle(0.3027)$\tabularnewline
 &  &  &  &  &  &  &  & $\arrowvert H-2\rightarrow L+1\rangle(0.3027)$\tabularnewline
 &  &  &  &  &  &  &  & \tabularnewline
VII\textsubscript{$xy$} &  & \textsuperscript{1}A &  & 5.75 &  & 1.905 &  & $\arrowvert H-2\rightarrow L+2\rangle(0.3136)$\tabularnewline
 &  &  &  &  &  &  &  & $\arrowvert H\rightarrow L;H-1\rightarrow L\rangle(0.2754)$\tabularnewline
 &  &  &  &  &  &  &  & \tabularnewline
VIII\textsubscript{$xy$} &  & \textsuperscript{1}A &  & 6.59 &  & 3.736 &  & $\arrowvert H-2\rightarrow L+6\rangle(0.2730)$\tabularnewline
 &  &  &  &  &  &  &  & $\arrowvert H-6\rightarrow L+2\rangle(0.2730)$\tabularnewline
 &  &  &  &  &  &  &  & \tabularnewline
IX\textsubscript{$xy$} &  & \textsuperscript{1}A &  & 7.11 &  & 3.910 &  & $\arrowvert H-5\rightarrow L+5\rangle(0.2417)$\tabularnewline
 &  &  &  &  &  &  &  & $\arrowvert H-1\rightarrow L;H-1\rightarrow L+1\rangle(0.2375)$\tabularnewline
\hline 
\end{tabular}
\end{table}